\documentclass[a4paper,12pt]{article}
\pdfoutput=1
\usepackage{graphicx,subfigure,amsmath,amssymb,color}
\usepackage{cite}

\newlength{\dinwidth}
\newlength{\dinmargin}
 \def\tb{\textcolor{black}}
  
 \def\la{ \langle}
  \def\ra{ \rangle}
 \def\r{ \gamma}
 \def\u{\mu}

 \def\lbd{\lambda}
 \def \d {{\rm d}}
  \def \k {\kappa}
  \def\eslash{\rlap{\hspace{0.02cm}/}{\epsilon}}
\begin{document}

\title{\bf  
Application of the light-front holographic wavefunction for heavy-light pseudoscalar meson in $B_{d,s}\to D_{d,s}P$ decays  
}
\author{Qin Chang$^{a,b}$, Shuai Xu$^{a}$ and Lingxin Chen$^{a}$
\\
{ $^a$\small Institute of Particle and Nuclear Physics, Henan Normal University, Henan 453007,  China}\\
{ $^b$\small Institute of Particle Physics, Central China Normal University, Wuhan 430079, China}\\
}
\date{}
 \maketitle

\begin{abstract}\noindent
In this paper we extend our analyses of the decay constant and distribution amplitude with \tb{an} improved holographic wavefunction to the heavy-light pseudoscalar mesons. \tb{In the evaluations, the  helicity-dependence of the holographic wavefunction is considered; and an independent mass scale parameter  
is employed to moderate the strong suppression induced by the heavy quark.    } Under the constraints from \tb{decay constants and masses of pseudoscalar mesons}, the $\chi^2$-analyses for the holographic parameters exhibit \tb{a rough consistence} with the results obtained by fitting the Regge trajectory. 
With the fitted parameters, the results for the decay constants and distribution amplitudes are presented. We then show their application in evaluating the $B_{d,s}\to D_{d,s}P$ decays,  in which the power-suppressed spectator scattering and weak annihilation corrections are first estimated. Numerically, the spectator scattering and weak annihilation corrections present a negative shift of about \tb{$ 0.7\%$} on the branching fractions;
while, the predictions
are still larger than the experimental data. Such small negative shift confirms the estimation based on the power counting rules. 
\end{abstract}


\newpage

\section{Introduction}
In recent years a semiclassical first approximation to strongly coupled QCD, light-front~(LF) holographic AdS/QCD,  has been developed~\cite{deTeramond:2005su,Brodsky:2006uqa,Brodsky:2007hb, Brodsky:2008pg,Brodsky:2008pf,deTeramond:2008ht,deTeramond:2010ge,Brodsky:2014yha} and successfully used  to predict  the spectroscopy of hadrons, dynamical observables such as form factors and structure functions, and the behavior of the running coupling in the nonperturbative domain {\it etc.}.  In this approach, the LF dynamics depend only on the boost invariant variables (the invariant mass \tb{$m_0$} or the invariant radial variable $\zeta$),  and the dynamical properties are encoded in the hadronic LF wavefunction  (LFWF), which has the form:
\begin{equation}
\psi(x,\zeta,\varphi)=e^{iL\varphi}X(x)\frac{\phi(\zeta)}{\sqrt{2\pi\zeta}}\,.
\end{equation}
The LF eigenvalue equation, $P_{\mu}P^{\mu}|\psi \ra=M^2|\psi \ra$, can be then reduced to an effective single-variable LF Schr\"odinger equation for $\phi(\zeta)$~\cite{deTeramond:2008ht},
\begin{equation}\label{eq:scheq}
\left(-\frac{{\rm d}^2}{{\rm d}\zeta^2 }-\frac{1-4L^2}{4\zeta^2}+U(\zeta)\right) \phi(\zeta)=M^2\phi(\zeta)\,,
\end{equation}
which is relativistic, frame independent and analytically tractable.

The the effective potential $U(\zeta)$ in Eq.~\eqref{eq:scheq}, which acts on the valence states and enforces confinement  at some scale,  is holographically related to a unique dilation profile in  anti-de Sitter~(AdS) space~\cite{deTeramond:2010ge,Brodsky:2014yha}. As a result, one arrives at a concise form of a color-confining harmonic oscillator after the holographical mapping,  $U(\zeta,J)=\lambda^2\zeta^2+2\lambda(J-1)$. Using this confining  potential, one can obtain the eigenvalues, which are the squares of the hadron masses,  by resolving the LF Schr\"odinger equation. In Refs.~\cite{deTeramond:2014asa,Dosch:2015nwa,Brodsky:2016yod,Brodsky:2016rvj}, with only one parameter \tb{$\lambda=\k^2$}, the observed light meson and baryon spectra are successfully described by extending superconformal quantum mechanics to the light-front and its embedding in AdS space. Moreover, very recently, the analyses are further extended to the heavy-light hadron family~\cite{Dosch:2016zdv}. 

The eigensolution of Eq.~(\ref{eq:scheq}) provides the holographic LFWF, which encodes the dynamical properties and is explicitly written as ~\cite{Brodsky:2007hb,Brodsky:2008pg}
\begin{equation} \label{eq:LFWFb}
\psi_{n,L}^{(0)}=\frac{1}{N}e^{iL\varphi}\sqrt{x(1-x)}\zeta^LL^L_n(|\lambda|\zeta^2)e^{-|\lambda|\zeta^2/2},
\end{equation}
for meson, where $N=\sqrt{(n+L)!/(n!\pi)}|\lambda|^{(L+1)/2}$ with LF angular momentum $L$ and radial excitation number $n$,  $\zeta^2=x(1-x){\bf b}_{\bot}^2$ with the invariant transverse impact variable ${\bf b}_{\bot}$ and momentum fraction $x$, and $L^L_n$ are the Laguerre Polynomials. 
\tb{
The holographic WF in the $\mathbf{k}_{\bot}$ space can be obtained via Fourier transform~\cite{Brodsky:2007hb,Brodsky:2014yha}. For the ground state, it is written as
 \begin{eqnarray}\label{eq:LFWFkT1}
\psi(z,\mathbf{k}_{\bot})=\,\frac{4\pi}{\kappa}\frac{1}{ \sqrt{z(1-z)}}\,e^{-\frac{\mathbf{k}_{\bot}^2}{2\kappa^2\,z(1-z)}}\,.
 \end{eqnarray}
}
This  holographic LFWF has been widely used to evaluate the hadronic observables, for instance, the decay constant, form factor and distribution amplitude~(DA) {\it etc.}~\cite{Brodsky:2007hb,Branz:2010ub,Brodsky:2011yv,Hwang:2012xf,Braga:2015jca,Vega:2009zb}.

\tb{It should be noted that the quark masses are not included in this  holographic LFWF,  and the helicity indices has been suppressed which is legitimate if the helicity dependence decouples from the dynamics.  For the phenomenological application, it is essential to restore both the quark mass and the helicity dependence of holographic LFWF. }

\tb{A simple generalization of the holographic WF for massive quarks follows from the assumption that the momentum space holographic WF is a function of the invariant off-energy shell quantity~\cite{Brodsky:2008pg,Brodsky:2014yha}, which implies the replacement in Eq.~\eqref{eq:LFWFkT1} that~\cite{Brodsky:2008pg,Brodsky:2014yha}
 \begin{eqnarray}\label{eq:replac}
\frac{\mathbf{k}_{\bot}^2}{\,z(1-z)}\,\to\,m_0=\frac{\mathbf{k}_{\bot}^2}{\,z(1-z)}+\Delta m^2\,,\quad \Delta m^2=\frac{m_q^2}{z}+\frac{m_{\bar{q}}^2}{1-z}\,.
 \end{eqnarray}
 }
Recently, it has been shown that the nonzero quark mass improves the description of data for $P$-to-photon ($P=\pi$ and $\eta^{(\prime)}$) transition form factors~\cite{Swarnkar:2015osa}. Unfortunately, for the heavy-light meson, the momentum fraction of light (anti-)quark  is pushed to a very small value, and the decay constant is strongly suppressed by the heavy quark mass~\cite{Dosch:2016zdv}. In order to remedy \tb{such} suppression, the mass term \tb{in the wavefunction} is further modified through \tb{the replacement~\cite{Dosch:2016zdv}
 \begin{eqnarray}
e^{-\frac{1}{2\k^2}\frac{m_q^2}{z}}\quad\rightarrow\quad e^{-\frac{\alpha^2}{2\k^2}\frac{m_q^2}{z}} 
 \end{eqnarray} 
($q$ is the heavy quark)} by introducing a scale factor $\alpha$, in which, the value $\alpha=1/2$ for the heavy-light meson is suggested. 
\tb{More generally, as suggested in Refs.~\cite{Branz:2010ub,Lyubovitskij:2010zz,Vega:2011ck,Gutsche:2014oua,Teramond:GHP2009}, the exponential term relevant to quark mass  can be absorbed in the longitudinal mode, $f(x,m_1,m_2)$; and the mass scale parameter, namely $\eta$, entering in $f(x,m_1,m_2)$ may not necessarily be identified with $\kappa$ characterizing the dilation field. The large values of $\eta$ are suggested to fit the spectra  and decay constants of heavy-light states~\cite{Branz:2010ub}.
}

The helicity dependence of holographic LFWF could be restored by introducing the helicity-dependent wavefunction $S_{h,\bar{h}}$. For the vector meson,  in analogy with the lowest order helicity structure of the photon LFWF in QED, the $S_{h,\bar{h}}$ with spinor structure, $\bar{u}_{h}\eslash^{\lbd}v_{\bar{h}}$ is introduced~\cite{Dosch:1996ss}, and  has been successfully used to describe diffractive $\rho$ meson electroproduction at HERA~\cite{Forshaw:2012im}. It is also used to predict the light-front distribution amplitudes~(LFDAs) of the $\rho$ and $K^*$ vector mesons~\cite{Ahmady:2012dy,Ahmady:2013cva}, the $B\to\rho\,,K^*$ form factors~\cite{Ahmady:2013cga,Ahmady:2014sva}, and further applied to rare $B \to K^* \mu^+ \mu^-$  and $B\to \rho \ell\bar{\nu}_{\ell}$ decays~\cite{Ahmady:2014cpa,Ahmady:2015gva,Ahmady:2015yea}. For the light pseudoscalar meson,  the helicity-dependent holographic LFWF is also studied very recently~\cite{Ahmady:2016ufq,Chang:2017}, and then confronts with a number of sensitive hadronic observables including the decay constants, DAs and $\xi$-moments of $\pi$ and $K$ mesons, the pion-to-photon transition form factor, and pure annihilation  $B_d\to K^+K^-$ and $B_s\to \pi^+ \pi^-$ decays. 

In this paper, we will extend the analyses of the improved holographic LFWF
to the heavy-light pseudoscalar meson. In the evaluations, the decay constant and holographic DA for the heavy-light pseudoscalar meson will be predicted. Then, in order to further test these results, we will apply them to evaluate the $B_{d,s}\to D_{d,s} P$ ($P=\pi$ and $K$) decays. These decay modes are dominated by the color-allowed tree topology, and have been evaluated, for instance, in the frameworks of the factorization assisted topological amplitude approach~\cite{Zhou:2015jba,Lu:2016uwv},  the naive factorization~(NF) with final state interaction~\cite{Chua:2007qw}, the QCD factorization~(QCDF)~\cite{Beneke2,Huber:2016xod} and the perturbative QCD~(pQCD)~\cite{Zou:2009zza,Rui:2011qc}. In the QCDF approach, the vertex QCD corrections at the levels of next-to-leading order~(NLO)~\cite{Beneke2} and next-to-next-to-leading order~(NNLO)~\cite{Huber:2016xod} have been evaluated in recent years. In this paper, besides of  the vertex amplitude, the spectator scattering and weak annihilation contributions will be evaluated  even though they are generally expected to be small based on the analysis of power counting~\cite{Beneke2}. 

Our paper is organized as follows. In section 2, the decay constants and DAs with the improved holographic LFWF for the heavy-light pseudoscalar mesons are studied.  With the hadronic observables obtained in section 2 as inputs, we further evaluate the $B_{d,s}\to D_{d,s} P$ decays in section 3.  Finally, we give our summary in section 4.

\section{Decay constant and distribution amplitude for  heavy-light pseudoscalar meson}

The general expression for the holographic LFWF with the inclusion of the helicity-dependence in $\mathbf{k}_{\bot}$ space can be written as 
\tb{
\begin{eqnarray}
\label{eq:LFWFP2}
\Psi_{h,\bar{h}}(z,\mathbf{k}_{\bot})=
S_{h,\bar{h}}(z,\mathbf{k}_{\bot}) \psi(z,\mathbf{k}_{\bot}) \,,
\end{eqnarray}
}where $h$($\bar{h}$) are the helicities of the (anti-)quark; $\psi(z,\mathbf{k}_{\bot})$ and $S_{h,\bar{h}}(z,\mathbf{k}_{\bot})$ are the radial and helicity-dependent  wavefunctions, respectively. 

\tb{
Following the proposal presented in Refs.~\cite{Teramond:GHP2009,Branz:2010ub,Lyubovitskij:2010zz,Vega:2011ck,Gutsche:2014oua}, a general form of the soft-wall holographic WF absorbing the quark mass term can be written as~\cite{Branz:2010ub} 
 \begin{eqnarray}\label{eq:modLFWF}
{\rm Eq.~\eqref{eq:LFWFkT1}}\quad\rightarrow \quad \psi(z,\mathbf{k}_{\bot})=\,\frac{4\pi}{\kappa}\frac{1}{ \sqrt{z(1-z)}}\,e^{-\frac{\mathbf{k}_{\bot}^2}{2\kappa^2\,z(1-z)}}\,f(z,m_q,m_{\bar{q}})\,,
 \end{eqnarray}
with the longitudinal mode 
 \begin{eqnarray}
f(z,m_q,m_{\bar{q}})\equiv N f(z) e^{-\frac{\Delta m^2 }{2\eta^2}}\,,
 \end{eqnarray}
 in which, $f(z)=1$, $\eta$ is the mass scale parameter, and $N$ is the normalization constant determined by
 \begin{eqnarray}\label{eq:NFact}
 \int_0^1\d z \,|f(z,m_q,m_{\bar{q}})|^2=1\,.
  \end{eqnarray}
It is noted that $\psi(z,\mathbf{k}_{\bot})$ given by Eq.~\eqref{eq:modLFWF} with $N$ obtained through Eq.~\eqref{eq:NFact} can automatically satisfy the normalization condition
\begin{eqnarray}\label{eq:norpsi}
 \int \frac { \d z\,\d^2\mathbf{k}_{\bot}}{2(2\pi)^3}|\psi(z,\mathbf{k}_{\bot})|^2=1\,,
\end{eqnarray}
which usually appears in literatures. }

\tb{For the dimensional parameter $\eta$ entering in the longitudinal mode,  the simplification $\eta=\k$  is usually  used in the studies for the light hadrons, even though it is not necessary as mentioned in the introduction and discussed in Refs.~\cite{Teramond:GHP2009,Branz:2010ub,Lyubovitskij:2010zz,Vega:2011ck}. It has been noted that such simplification leads to the very strong suppression on the decay constants for the heavy-light mesons~\cite{Dosch:2016zdv,Branz:2010ub}.  In order to remedy such suppression, the mass term of heavy quark in the wavefunction is rescaled through $m_q^2\to \alpha^2m_q^2$ with $\alpha\sim 1/2$ in Ref.~\cite{Dosch:2016zdv}. Alternatively, we adopt the proposal presented in Refs.~\cite{Branz:2010ub,Lyubovitskij:2010zz,Vega:2011ck}, and assume that $\eta=\k$ only in the limit of massless quark,  $m_{q,\bar{q}}\to 0$, and $\eta>\k$ for the other cases. Numerically, we take $\eta/\k=1\,,1.5\,,2.5$ for $q=u(d),s,c(b)$~($q$ is the relatively heavy quark in a meson) for simplicity.  Here, we would like to clarify that such strategy used in this paper is a phenomenological approach to remedy the strong suppression effect caused by the heavy quark mass, and more theoretical efforts are required to explore the underlying mechanism.
}

In the recent works, the helicity-dependent wavefunctions for the pseudoscalar meson with spinor structures, $\bar{u}_{h}(i\r_5)v_{\bar{h}}$ and $\bar{u}_{h}(i\frac{m_0}{2p^+}\r^+\r_5+i\r_5)v_{\bar{h}}$, are introduced and then confronted with hadronic observables, such as the decay constants of $\pi$ and $K$ mesons, their $\xi$-moments, the pion-to-photon transition form factor and the $\bar{B}_{s}\to \pi^+ \pi^-$ and $\bar{B}_{d}\to K^+K^-$ pure annihilation  decays~\cite{Ahmady:2016ufq,Chang:2017}.  
\tb{The helicity-dependent wavefunction, $S_{h,\bar{h}}(z,\mathbf{k}_{\bot})$, can be obtained by the interaction-independent Melosh transformation~\cite{Melosh:1974cu}. Explicitly, the covariant form  of $S_{h,\bar{h}}(x,\mathbf{k}_{\bot})$ is written as~\cite{Jaus:1989au,Hwang:2010hw}
\begin{eqnarray}\label{eq:wfS}
S_{h,\bar{h}}(z,\mathbf{k}_{\bot})=\frac{\bar{u}_{h}(k_1)\,i\gamma_5\, v_{\bar{h}}(k_2)}{\sqrt{2}\sqrt{m_0^2-(m_q-m_{\bar{q}})^2}}\,,
\end{eqnarray}
which satisfies the normalization condition
\begin{eqnarray}\label{eq:SNor}
\sum_{h,\,\bar{h}}S_{h,\bar{h}}^{\dagger}(z,\mathbf{k}_{\bot})S_{h,\bar{h}}(z,\mathbf{k}_{\bot})=1\,.
\end{eqnarray} 
It is noted that the total normalization condition
\begin{eqnarray}\label{eq:nortot}
\sum_{h,\bar{h}}\int \d z \frac{\d^2\mathbf{k}_{\bot}}{2(2\pi)^3} |\Psi_{h,\bar{h}}(z,\mathbf{k}_{\bot})|^2=1\,
\end{eqnarray}
is also  automatically satisfied by the improved holographic LFWF $\Psi_{h,\bar{h}}(z,\mathbf{k}_{\bot})$ given by Eq.~\eqref{eq:LFWFP2} with $\psi(z,\mathbf{k}_{\bot})$ and $S_{h,\bar{h}}(z,\mathbf{k}_{\bot})$ given respectively by Eqs.~\eqref{eq:modLFWF} and  \eqref{eq:wfS}.
}

The decay constant of pseudoscalar meson is defined by 
\begin{eqnarray}
\label{eq:dc}
\la 0 | \bar q\r_\u\r_5 q|P(p)\ra=if_P p_\u \,.
\end{eqnarray}
In the framework of LF quantization, with the Lepage-Brodsky (LB) conventions and light-front gauge~\cite{Brodsky:1997de,Lepage:1980fj}, a hadronic eigenstate $|P\ra$ can be expanded in a complete Fock-state basis of noninteracting 2-particle states as
\begin{equation}
|P\ra = \sum_{h,\bar{h}} \int \frac{\d k^+ \d^2{\bf k_{\bot}}}{(2\pi)^32\sqrt{k^+(p^+-k^+)}} \Psi_{h,\bar{h}}\left(k^+/p^+,{\bf k}_{\bot}\right)|k^+,k_{\bot},h;p^+-k^+,-k_{\bot},\bar{h}\ra \,.
\label{eq:Fockexp}
\end{equation}
With the LF helicity spinors $ u_h$ and $v_{h}$, the Dirac (quark) field is expanded as
\begin{equation}
\psi_+(x)= \int \frac{\d k^+}{\sqrt{2k^+}}\frac{ \d^2{\bf k}_{\bot}}{(2\pi)^3} \sum_h [b_{h} (k) u_h(k)e^{-ik\cdot x} + d^{\dagger}_{h} (k) v_{h} (k)e^{ik\cdot x}] \,,
\label{eq:qfexp}
\end{equation}
 in terms of particle creation and annihilation operators, which satisfy the equal LF-time anti-commutation relations
\begin{equation}
\{b^{\dagger}_{h} (k), b_{h'} (k^{\prime}) \}=\{d^{\dagger}_{h} (k), d_{h'} (k^{\prime}) \}= (2\pi)^3  \delta(k^+-k'^{+})\delta^2({\bf k}_{\bot}-{\bf k}'_{\bot}) \delta_{h h'}.
\label{anticommutation}
\end{equation}
Using above formulae, the left-hand-side of Eq.~(\ref{eq:dc}) for $\mu=+$ can be expressed as
\begin{eqnarray}
 \la 0 | \bar{q} \r^+\r_5 q|P(p) \ra&=&
\sqrt{N_c} \sum_{h,\bar{h}} \int \frac{\d z\d^2{\bf k}_{\bot}}
{(2\pi)^32\sqrt{z\bar{z}}}
\Psi_{h,\bar{h}}(z,{\bf k}_{\bot})
\bar{v}_{\bar{h}}(\bar{z},-{\bf k}_{\bot}) \r^+\r_5 u_h(z,{\bf k}_{\bot})\;.
\label{eq:meder1}
\end{eqnarray}
Then, further using Eqs.~\eqref{eq:LFWFP2}, \eqref{eq:wfS} and \eqref{eq:dc}, and $ \bar{v}_{\bar{h}} \r^+\r_5 u_h=\pm\,  2{\sqrt{z\bar{z}}} p^+ \delta_{h\pm,\bar{h}\mp}$, we finally arrive at  
\tb{
\begin{eqnarray}
f_P=\frac{\sqrt{N_c}}{\pi}\int_0^1\d z\int\frac{\d^2{\bf k}_{\bot}}{(2\pi)^2}\frac{ \psi(z,{\bf k}_{\bot})}{\sqrt{z\bar{z}}}\frac{\left(\bar{z}m_q+zm_{\bar{q}}\right)}{\sqrt{2}\sqrt{m_0^2-(m_q-m_{\bar{q}})^2}}\,.
\end{eqnarray}
}

Using the theoretical formulae given above, we then present our numerical evaluation. In the framework of LF holographic QCD, the basic inputs include the mass scale parameter $\kappa$ and the effective quark masses. 
The effective quark masses, in principle, should be universal in a specific theoretical framework of holographic QCD; \tb{while, the $\kappa$ is generally different for various $(q,\bar{q}^{\prime})$ states.} 

For the light hadrons, the values of holographic QCD parameters have been well determined in the past years. The value \tb{$\kappa\sim [0.5,0.63]\,{\rm GeV}$} is generally expected by extracting from hadronic observables, for instance, Refs.~\cite{Ahmady:2013cva,Brodsky:2014yha,Brodsky:2016rvj,Deur:2016opc,Prosperi:2006hx,Forshaw:2012im,Ahmady:2016ujw}, in which \tb{$\k=0.59\,{\rm GeV}$} for light pseudoscalar mesons is suggested by fitting the light-quark spectrum~\cite{Brodsky:2014yha}. 
The light-quark mass $m_{u,d}=46\,{\rm MeV}$ and $m_s=357 \,{\rm MeV}$~\cite{Brodsky:2014yha} are obtained by fitting to $m_{\pi,K}$ \tb{with only the soft-wall potential considered. In some works for evaluating the hadronic observables, the constituent mass $m_{u,d}\sim 350\,{\rm MeV}$ and $m_s\sim480 \,{\rm MeV}$ are used, for instance, Ref.~\cite{Ahmady:2013cva}. In addition, after considering the color Coulomb-like potential, the much larger values $m_{u,d}\sim 420\,{\rm MeV}$ and $m_s\sim570 \,{\rm MeV}$ are suggested~\cite{Branz:2010ub}. }

\tb{
For the heavy-light hadrons, the values of holographic parameters are not  determined precisely for now.  Very recently, fitting to the masses of the ground states for both mesons and baryons containing a charm quark or a bottom quark, the best-fit results $m_c=1.327~(1.547)\,{\rm GeV}$, $m_b=4.572\,( 4.922)\,{\rm GeV}\,$ with $\alpha=0.5\,(1)$ are obtained~\cite{Dosch:2016zdv}. In addition, the best-fit values $\k_{\bar{q}c\,,\bar{s}c}$ $\sim$ $[0.655,0.736]\,{\rm GeV}\,,[0.735,0.766]\,{\rm GeV}$  and $\k_{\bar{q}b\,,\bar{s}b}$ $\sim$ $[0.963,1.13]\,{\rm GeV}\,,[1.11,1.16]\,{\rm GeV}$ ($q=u,d$)  have also been obtained by fitting the Regge slopes~\cite{Dosch:2016zdv}. The fitted values of $\k$ for the light and heavy-light mesons~\cite{Brodsky:2014yha,Dosch:2016zdv} are summarized in Table~\ref{tab:lamb}.
}

\begin{table}[t]
\caption{\label{tab:lamb} \tb{The fitted results (the third row) for  the mass scale parameter \tb{$\k$} in unit of ${\rm GeV}$ under the constraints  from the  decay constants and meson masses. The second row corresponds to the results obtained by fitting to the hadron spectra~\tb{\cite{Brodsky:2014yha,Dosch:2016zdv}}.}}
\renewcommand*{\arraystretch}{1.1}
\begin{center}\setlength{\tabcolsep}{5pt}
\begin{tabular}{lccccccccc}
\hline\hline
      & $\kappa_{\bar{q}q}$&$\kappa_{\bar{q}s}$&$\k_{\bar{q}c}$ & $\k_{\bar{s}c}$ &$\k_{\bar{q}b}$ & $\k_{\bar{s}b}$\\ \hline
Refs.~\cite{Dosch:2016zdv,Brodsky:2014yha}&$0.59$&$0.59$ &$0.655$ &$0.735$ &$0.963$&$1.110$\\
this work & $0.548^{+0.014}_{-0.012}$& $0.626^{+0.012}_{-0.014}$&$0.846^{+0.030}_{-0.030}$ &  $0.957^{+0.026}_{-0.028}$  &$1.067^{+0.028}_{-0.023}$ & $1.144^{+0.027}_{-0.018}$ \\
\hline\hline
\end{tabular}
\end{center}
\end{table}

The leptonic decay constants provide severe tests for the adequacy of the wavefunction and constraints on the spaces of holographic parameters.  For the charged heavy-light mesons, the decay constants can be  determined \tb{ experimentally} through the leptonic $P\to l\bar{\nu}$  decays. The updated world averaged experimental results~\cite{PDG} are summarized in the second row of Table~\ref{tab:dc}, and will be used in the following $\chi^2$ analyses~\footnote{The averaged lattice QCD~(LQCD) results, $f_{B_d}=(187.1  \pm 4.2){\rm MeV}$ and $f_{B_s}=(227.2 \pm 3.4){\rm MeV}$, are used in the fit because there is not available data for $f_{B_s}$. The former is in consistence with the data $(188  \pm 25){\rm MeV}$.}. 
\tb{For consistence, the fit to the light mesons is also revisited in this paper, and the data $f_{\pi}=(130.28\pm0.26){\rm MeV}$ and $f_{K}=(156.09\pm0.49){\rm MeV}$~\cite{PDG} are used. Besides the decay constants, the mesons' messes are also taken into account in our fits. The averaged data given by PDG~\cite{PDG} are used~\footnote{In our $\chi^2$-fit, as a conservative choice, an additional $1\%$ error is assigned to the experimental data of observable if its significance is larger than $100\sigma$ errors.}.  }

\tb{
In this paper, the contributions of an additional color Coulomb-like interaction, $V(r)=-4\alpha_s/3r$, induced by the one-gluon exchange~\cite{Sergeenko:1994ck,Gershtein:2006ng} are also included.  It can be achieved phenomenologically by extending $U\to U+U_C$, where $U_C$ is the contribution of color Coulomb-like potential, and has been studied in, for instance, Refs.~\cite{Branz:2010ub,Lyubovitskij:2010zz,Gutsche:2012ez,Gutsche:2014oua}. It is noted that the contribution of  $U_C$ to the mass $M^2$ is negative and proportional to the quark mass squared. 
 Such corrections can be included in the form of a constant term~\cite{Branz:2010ub,Gutsche:2012ez,Gutsche:2014oua}
\begin{eqnarray}
\Delta M_C^2=-\frac{64\alpha_s^2(\mu_{\bar{q}q}) m_qm_{\bar{q}}}{9(n+L+1)^2}\,,
\end{eqnarray}
with $\mu_{\bar{q}q}=2m_qm_{\bar{q}}/(m_q+m_{\bar{q}})$. The strong coupling $\alpha_s$ depend on the quark flavor and has the following ``freezing" form~\cite{Branz:2010ub,Badalian:2004xv,Ebert:2009ub} 
\begin{eqnarray}
\alpha_s(\mu^2)=\frac{12\pi}{(33-2N_f)\ln\frac{\u^2+M_B^2}{\Lambda^2}}\,,
\end{eqnarray}
where, $N_f$ is the number of flavors, $\Lambda$ is the QCD scale parameter, $M_B$ is the back ground mass. In the numerical evaluation, we take $\Lambda=415{\rm MeV}$ and $M_B=855{\rm MeV}$~\cite{Branz:2010ub,Badalian:2004xv}. Finally, the master formula for the mass of pseudoscalar meson~(ground-state) reads~\cite{Branz:2010ub}
\begin{eqnarray}
M^2=\int_0^1\d z\, \left(\frac{m_q^2}{z}+\frac{m_{\bar{q}}^2}{1-z}\right) \,f^2(z,m_q,m_{\bar{q}})+\Delta M_C^2\,.
\end{eqnarray}
}

\begin{table}[t]
\caption{\label{tab:dc} \tb{The experimental \tb{data} and theoretical \tb{results} for the decay constants of light and heavy-light mesons in unit of ${\rm MeV}$. See text for explanation.}}
\let\oldarraystretch=\arraystretch
\renewcommand*{\arraystretch}{1.1}
\begin{center}\setlength{\tabcolsep}{2pt}
\begin{tabular}{lcccccccccccc}
\hline\hline
   &$f_{D^+}$  &$  f_{D_s}$  & $f_{D_s}/f_{D^+}$ &   $f_{B^-}$ &  $f_{B_s}$   &$f_{B_s}/f_{B^-}$  \\  \hline
Exp.~\cite{PDG}        & $203.7\pm4.7$ & $257.8\pm4.1$   &$1.266\pm0.035$ &$188\pm25$ & ---   &---\\
LQCD~\cite{Aoki:2016frl}  &$211.9\pm1.1$ &$249.0\pm1.2$  &$1.173\pm0.003$  &$187.1\pm4.2$ &$227.2\pm3.4$&$1.215\pm0.007$   \\
QCDSR~\cite{Narison:2015nxh}   &$204.0\pm4.6$  &$243.2\pm4.9$ &$1.170\pm0.023$ &$204.0\pm5.1$ &$234.5\pm4.4$   &$1.154\pm0.021$ \\
LFQM~\cite{Hwang:2009qz}    &$205.8\pm8.9$ &$264.5\pm17.5$ &$1.29\pm0.07$ &$204\pm31$ &$270.0\pm42.8$ &$1.32\pm0.08$\\
LFHQCD~\cite{Dosch:2016zdv}  &$199\,(127)$   &$216\,(159)$  &$1.09\,(1.25)$  &$194\,(81)$ &$229\,(117)$ &$1.18\,(1.44)$\\
\tb{this work}  &$214.2^{+7.6}_{-7.8}$   &$253.5^{+6.6}_{-7.1}$  &$1.184^{+0.054}_{-0.052}$  &$191.7^{+7.9}_{-6.5}$ &$225.4^{+7.9}_{-5.3}$ &$1.176^{+0.056}_{-0.053}$\\
\hline\hline
\end{tabular}
\end{center}
\end{table}

\begin{figure}[t]
\begin{center}
\subfigure[]{\includegraphics[scale=0.42]{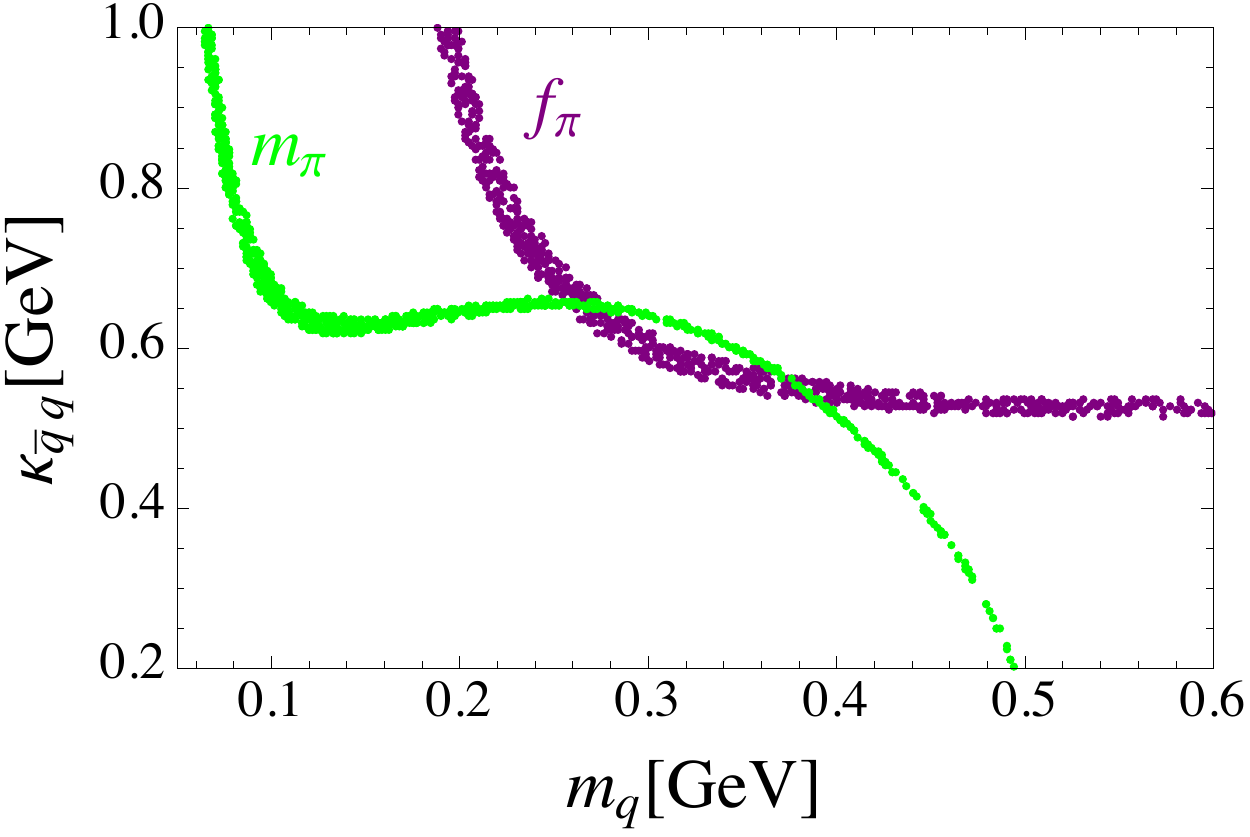}}~~
\subfigure[]{\includegraphics[scale=0.42]{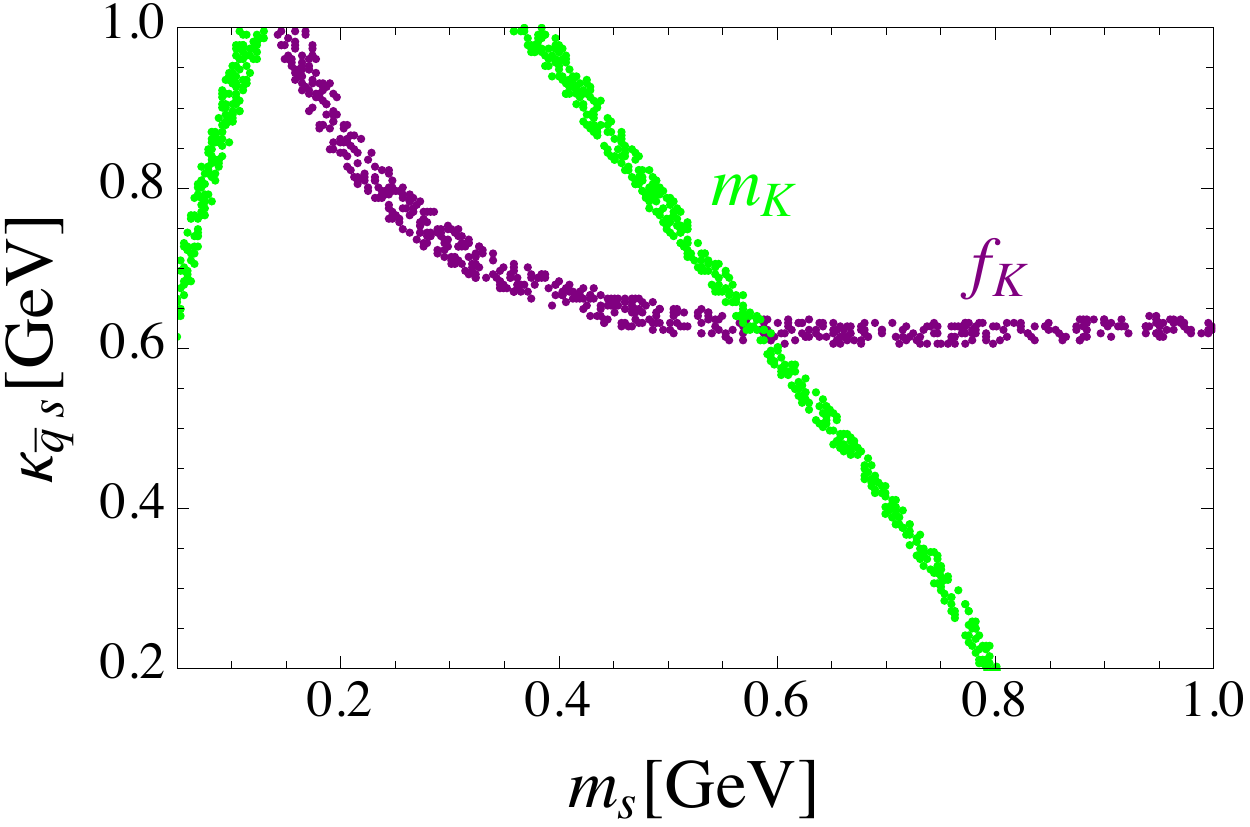}}~~
\subfigure[]{\includegraphics[scale=0.42]{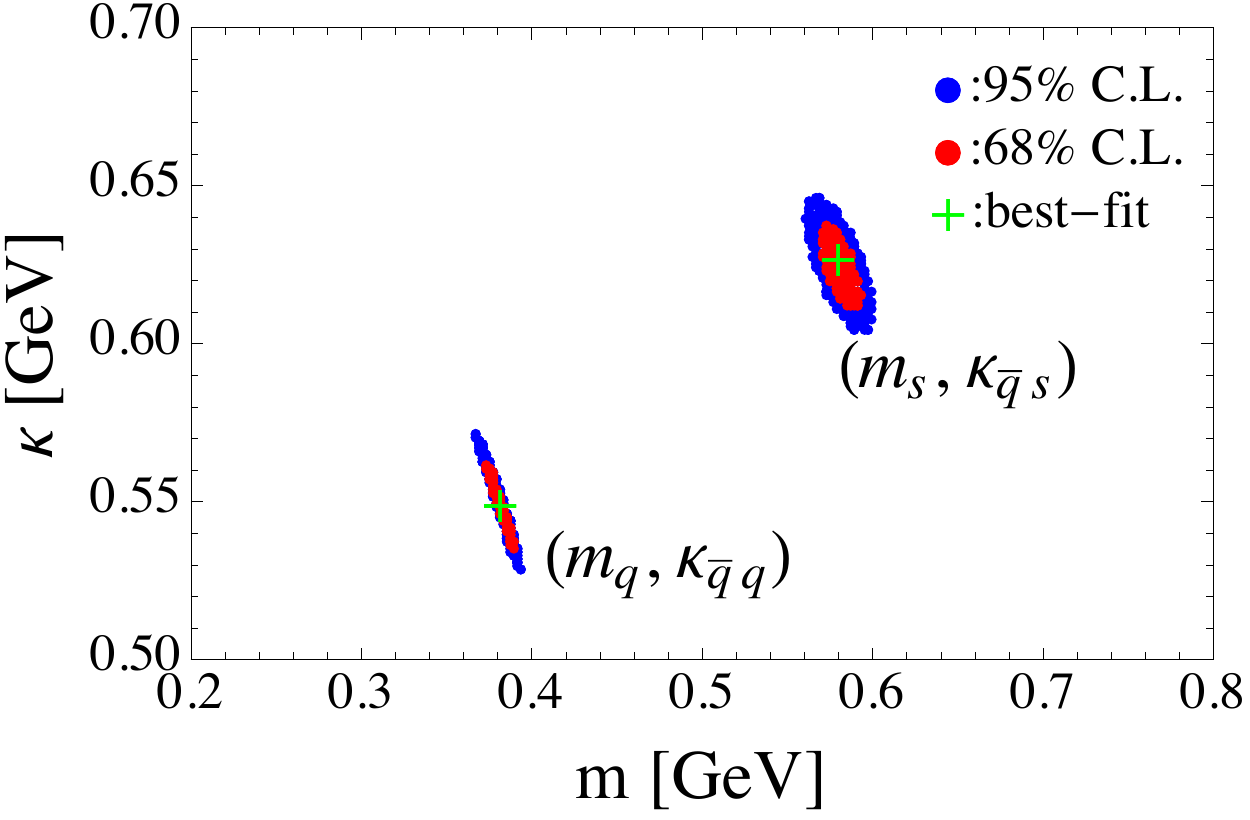}}
\caption{\label{fig:fit1}\tb{The fitted spaces for the holographic parameters $(\kappa_{\bar{q}q},m_q)$ and $(\kappa_{\bar{q}s},m_s)$ with $q=u,d$.  Figs. (a) and (b) show the results under the constraints from the decay constants and messes of mesons~($\pi$ and $K$), respectively, at $95\%$ C.L.; Fig. (c) shows the results under the combined constraints at $68\%$ and $95\%$ C.L..}}
\end{center}
\end{figure}

\tb{
Under the constraints form the decay constants and masses of $\pi$ and $K$ mesons, the allowed spaces of $(\kappa_{\bar{q}q},m_q)$ and $(\kappa_{\bar{q}s},m_s)$~($q=u,d$) are shown in Fig.~\ref{fig:fit1}. It can be clearly seen that the constraints on the parameter spaces are very strong due to the precisely measured data. The numerical results for  $\kappa$ are summarized in Table~\ref{tab:lamb}, in which the results obtained by fitting to the hadron spectra~\cite{Brodsky:2014yha,Dosch:2016zdv} are also listed for comparison. We find that: 
(i) For $(\kappa_{\bar{q}q},m_q)$, two solutions around $(0.65,0.28)\,{\rm GeV}$ and $(0.55,0.38)\,{\rm GeV}$ can be clearly seen from Fig.~\ref{fig:fit1}~(a), in which the later having relatively small $\kappa_{\bar{q}q}\sim 0.55 \,{\rm GeV}$ is employed in the following evaluations.  
(ii) As Fig.~\ref{fig:fit1}~(c) and Table~\ref{tab:lamb}  show, one can find $\kappa_{\bar{q}s}/\kappa_{\bar{q}q}\sim 1.14>1$, which is caused by the flavor symmetry-breaking effect exhibited by the data $f_{K}/f_{\pi}\sim 1.20$.  Our fitted results $\kappa_{\bar{q}q}\sim0.54\,{\rm GeV}$ and $\kappa_{\bar{q}s}\sim0.63\,{\rm GeV}$ are in consistence with the result $\kappa_{\bar{q}q,\bar{q}s}\sim0.59\,{\rm GeV}$~\cite{Brodsky:2014yha}. 
(iii) For the quark mass, we find that our results 
\begin{eqnarray}\label{eq:massqs}
m_{q}=0.382^{+0.007}_{-0.008} \,{\rm GeV}\,,\qquad m_s=0.580^{+0.012}_{-0.011}\,{\rm GeV}
\end{eqnarray}
at 68\% C.L. are significantly  larger than the ones $m_{q}=0.046 \,{\rm GeV}$ and  $m_s=0.357\,{\rm GeV}$~\cite{Brodsky:2014yha} because of the contribution of color Coulomb-like potential considered in this paper which leads to a negative shift of $M^2$.
}

\begin{figure}[t]
\begin{center}
\subfigure[]{\includegraphics[scale=0.42]{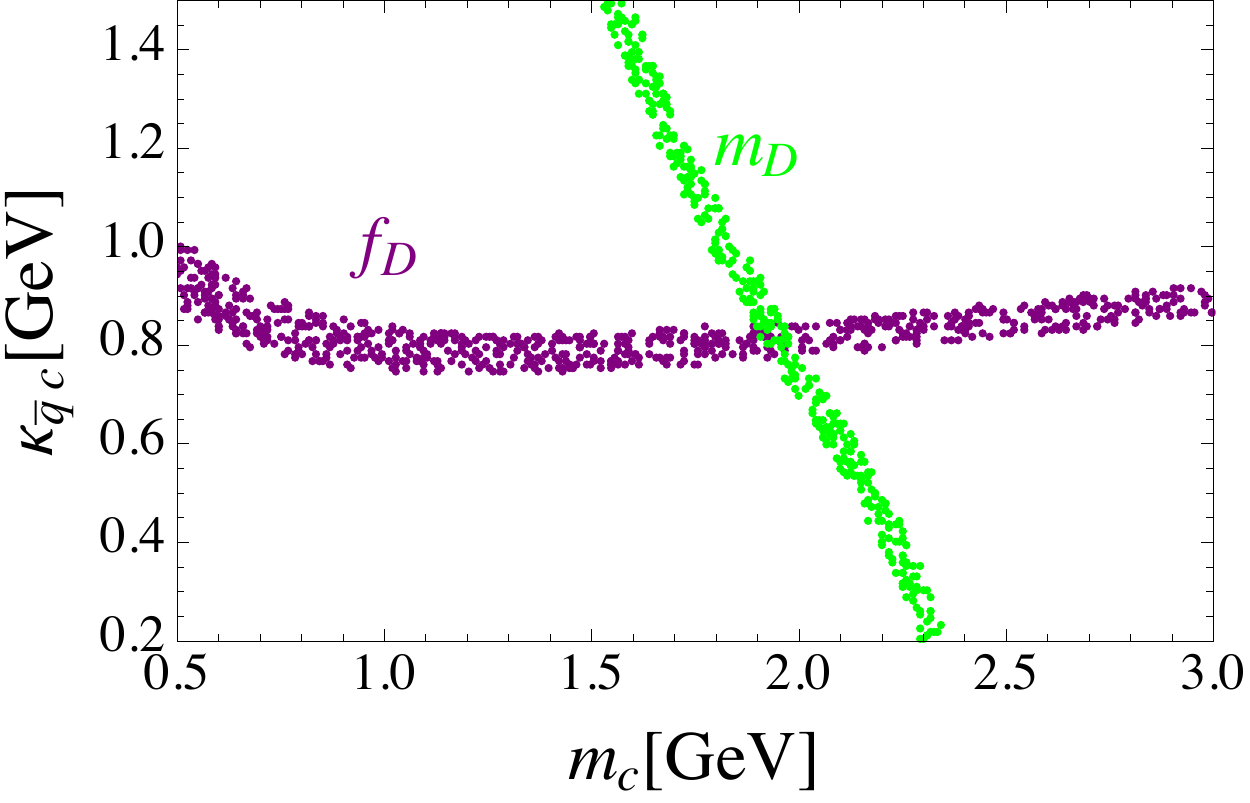}}~~
\subfigure[]{\includegraphics[scale=0.42]{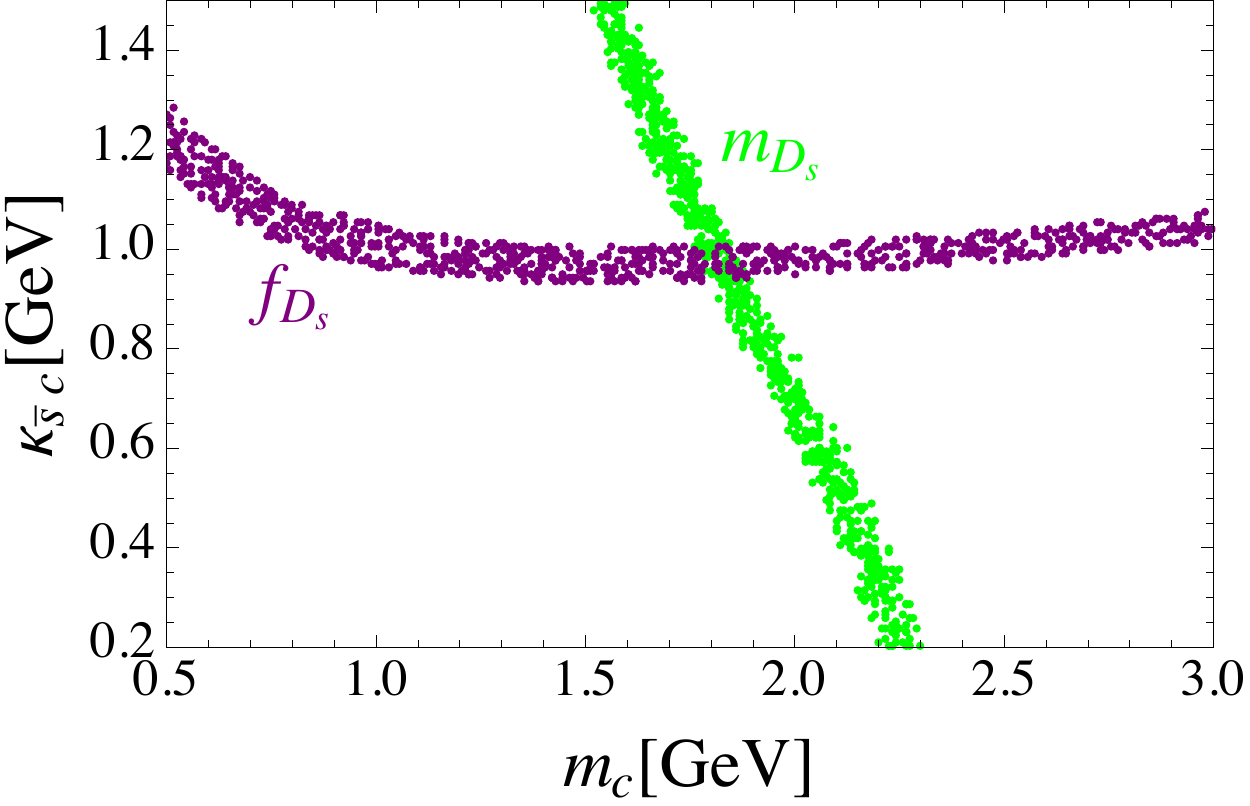}}~~
\subfigure[]{\includegraphics[scale=0.42]{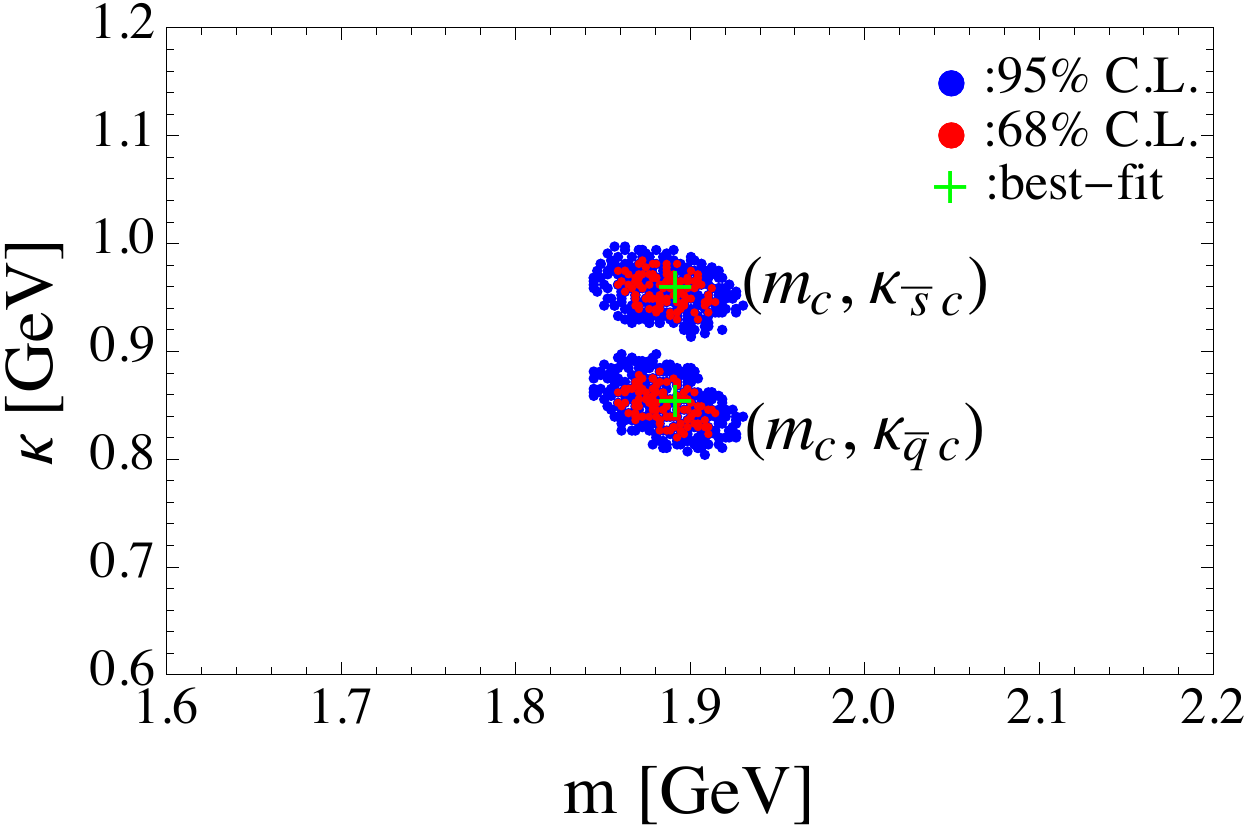}}
\caption{\label{fig:fit2} \tb{The fitted spaces for the holographic parameters $(\kappa_{\bar{q}c},m_c)$ and $(\kappa_{\bar{s}c},m_c)$ with best-fit values of $m_q$ and $m_s$ as inputs. The other captions are the same as Fig.~\ref{fig:fit1}.}}
\end{center}
\end{figure}

\begin{figure}[ht]
\begin{center}
\subfigure[]{\includegraphics[scale=0.42]{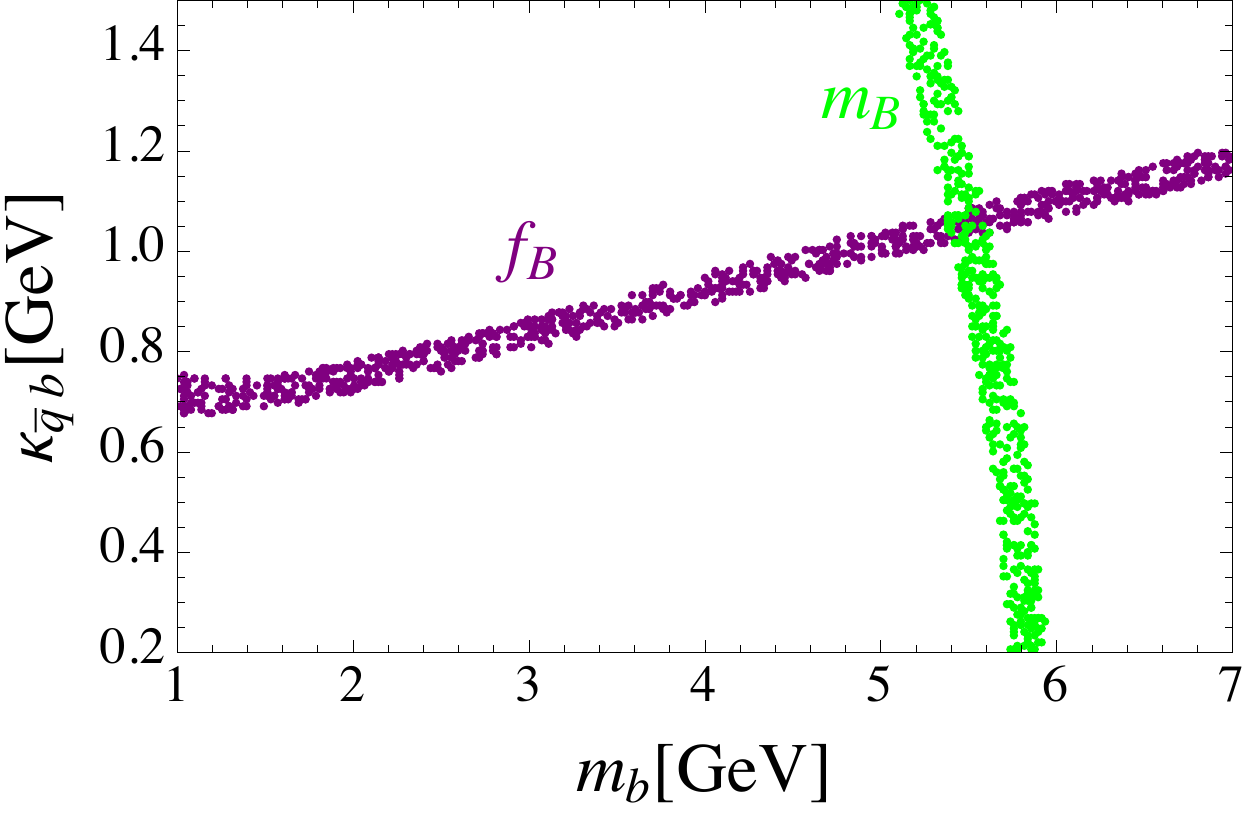}}~~
\subfigure[]{\includegraphics[scale=0.42]{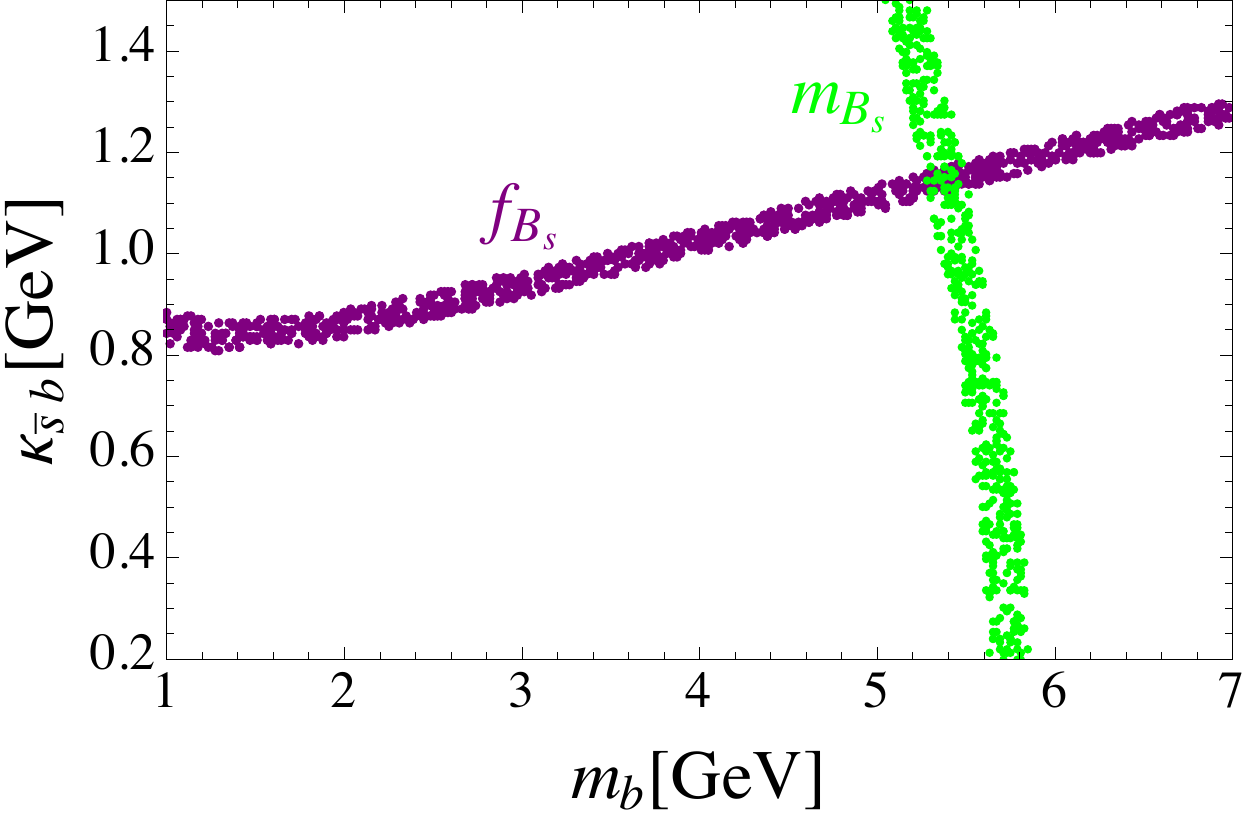}}~~
\subfigure[]{\includegraphics[scale=0.42]{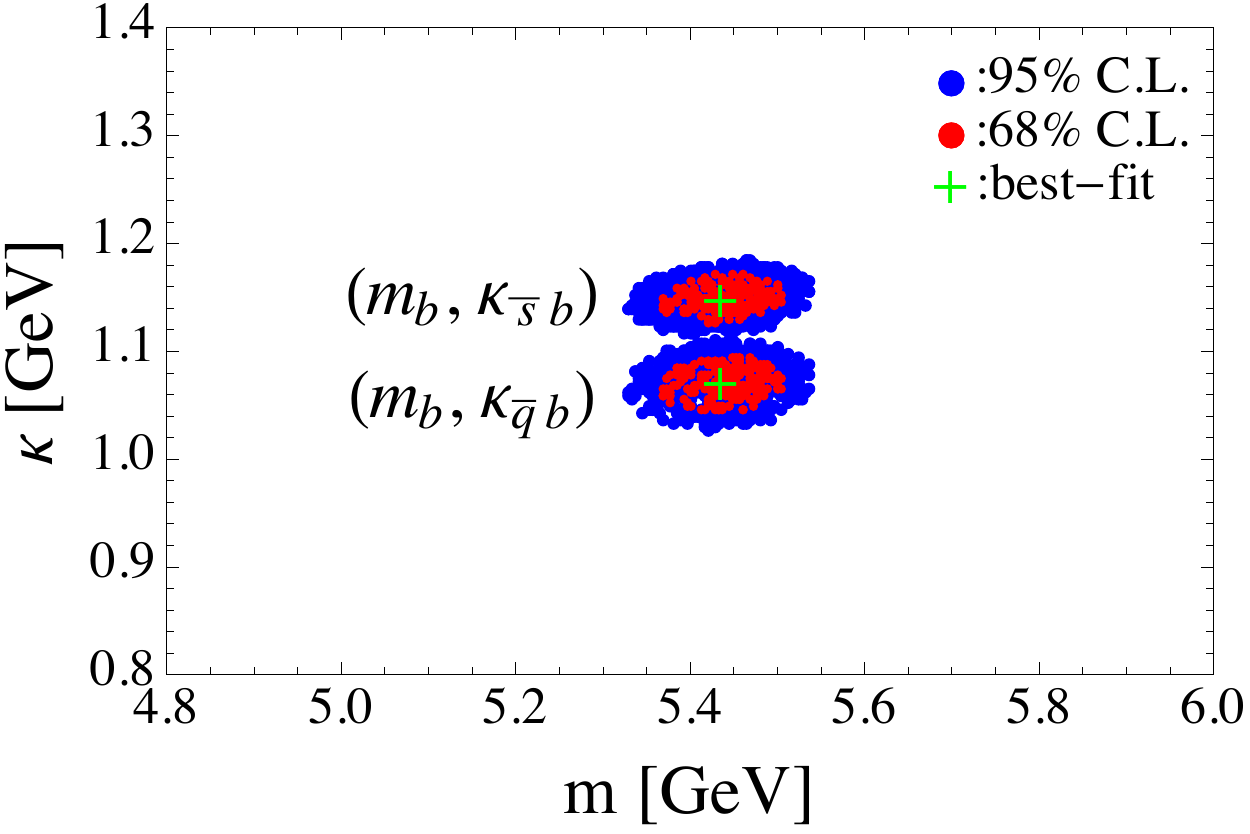}}
\caption{\label{fig:fit3} \tb{The fitted spaces for the holographic parameters $(\kappa_{\bar{q}b},m_b)$ and $(\kappa_{\bar{s}b},m_b)$ with best-fit values of $m_q$ and $m_s$ as inputs. The other captions are the same as Fig.~\ref{fig:fit1}.}}
\end{center}
\end{figure}

\tb{
With the best-fit values of light-quark masses given by Eq.~\eqref{eq:massqs} as inputs, the allowed spaces of the holographic parameters for the $D_{(s)}$ and $B_{(s)}$ mesons are shown by Figs.~\ref{fig:fit2} and \ref{fig:fit3}, respectively. The numerical results for  $\kappa$ are also summarized in Table~\ref{tab:lamb}. It can be clearly seen  that: 
(i) As Figs.~\ref{fig:fit2} (a) and (b) show, the mass $m_{D_{(s)}}$ dominates the constraint on the $c-$quark mass, and the parameter $\kappa_{\bar{q}c}\,(\kappa_{\bar{s}c})$ is mainly determined by the decay constant $f_{D_{(s)}}$; the case for  $B$ and $B_s$ mesons is similar which can be seen from  Figs.~\ref{fig:fit3} (a) and (b). Under the combined constraints, the holographic parameters and quark masses are strictly bounded. 
(ii)  Similar to the case for $\pi$ and $K$ system, one can find that $\kappa_{\bar{s}c}/\kappa_{\bar{q}c}\sim 1.13>1$ and $\kappa_{\bar{s}b}/\kappa_{\bar{q}b}\sim 1.07>1$  is required to provide sufficient flavor symmetry-breaking resource for fitting $f_{D_s}/f_{D}$ and $f_{B_s}/f_{B}$. Comparing with the best-fit values of $\kappa$ obtained by fitting to the Regge trajectories~\cite{Dosch:2016zdv}, one can clearly see from Table~\ref{tab:lamb} that our results for $\kappa_{\bar{q}b}$ and $\kappa_{\bar{s}b}$ agree with the ones in Ref.~\cite{Dosch:2016zdv}, but our results for $\kappa_{\bar{q}c}$ and $\kappa_{\bar{s}c}$ are relatively larger which is mainly caused by the constraints from $f_{D_{(s)}}$. 
(iii) Due to the negative contribution  to $M^2$ induced by color Coulomb-like potential, our results for the heavy quark masses are also larger than the ones in Ref.~\cite{Dosch:2016zdv}. Numerically, we obtain
\begin{eqnarray}\label{eq:masscb}
 m_c=1.882^{+0.025}_{-0.030}\,{\rm GeV}\,,\qquad 
 m_b=5.435^{+0.071}_{-0.067}\,{\rm GeV}\,
\end{eqnarray}
at 68\% C.L..}

With above fitted results as inputs, our \tb{theoretical results} for the decay constants are listed in the table~\ref{tab:dc}~\footnote{\tb{Our theoretical results here should be treated as ``posterior predictions'' since they are based on the inputs obtained by fitting current data.}}, in which the world averaged results based on the lattice QCD (LQCD)~\cite{PDG,Aoki:2016frl} with $N_f = 2+1(+1)$ and the QCD sum rules (SR)~\cite{Narison:2015nxh},  the theoretical predictions based on a light-front quark model (LFQM)~\cite{Hwang:2009qz} and the LF holographic QCD~(LFHQCD) with $\alpha=0.5\,(1)$~\cite{Dosch:2016zdv} are also listed for comparison.
\tb{From table~\ref{tab:dc}, it can be found that our results are generally in consistence with the data and the results in the other theoretical framework. Moreover, comparing with the previous results in Ref.~\cite{Dosch:2016zdv}, one can find that the results of holographic QCD can be improved when the  helicity-dependent wavefunction is taken into account. }

\begin{figure}[t]
\begin{center}
\subfigure[]{\includegraphics[scale=0.42]{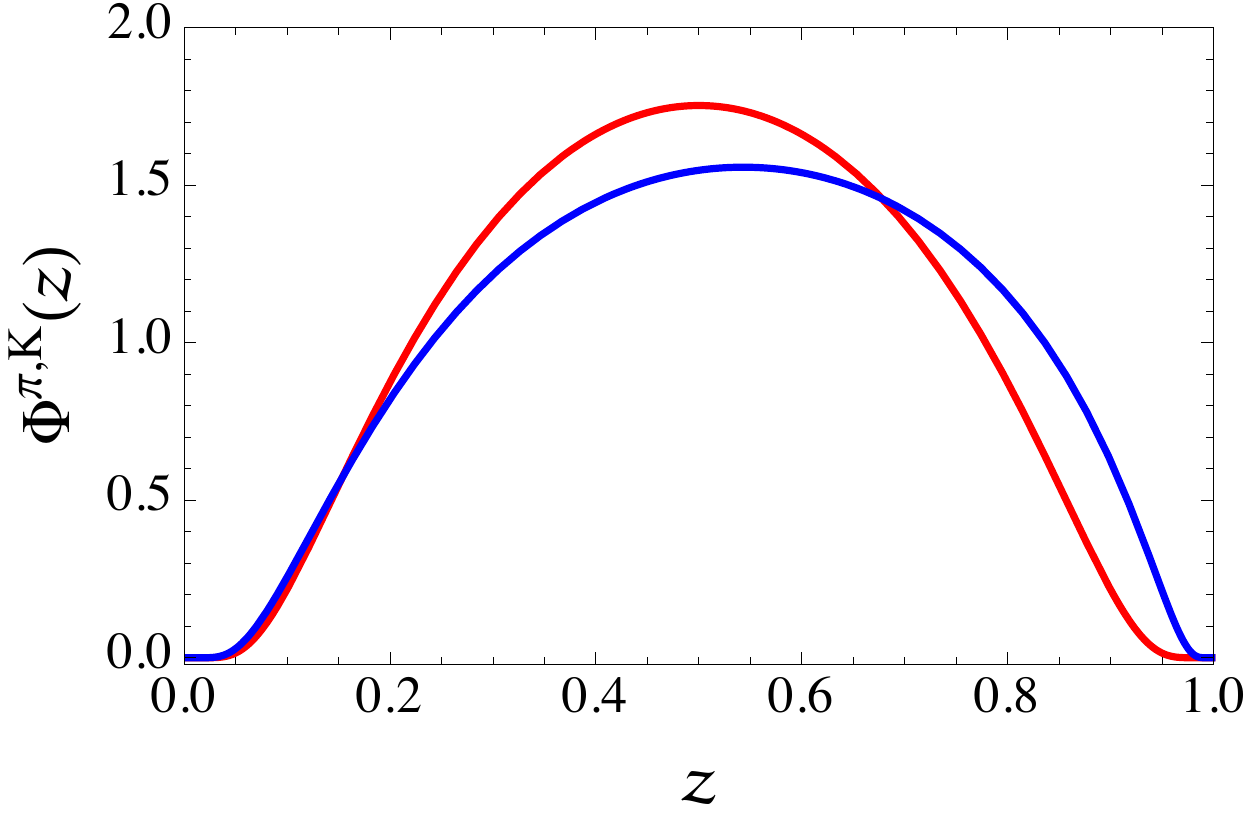}}~~
\subfigure[]{\includegraphics[scale=0.42]{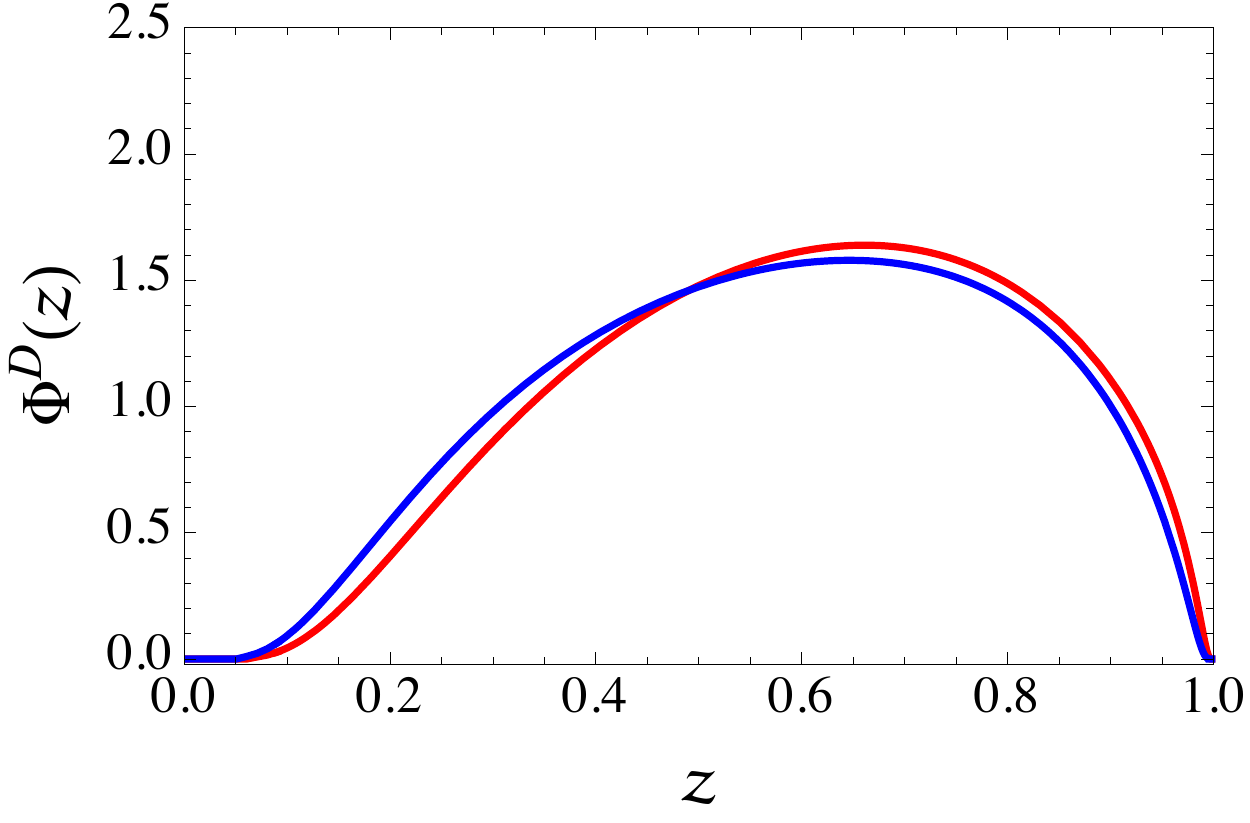}}~~
\subfigure[]{\includegraphics[scale=0.42]{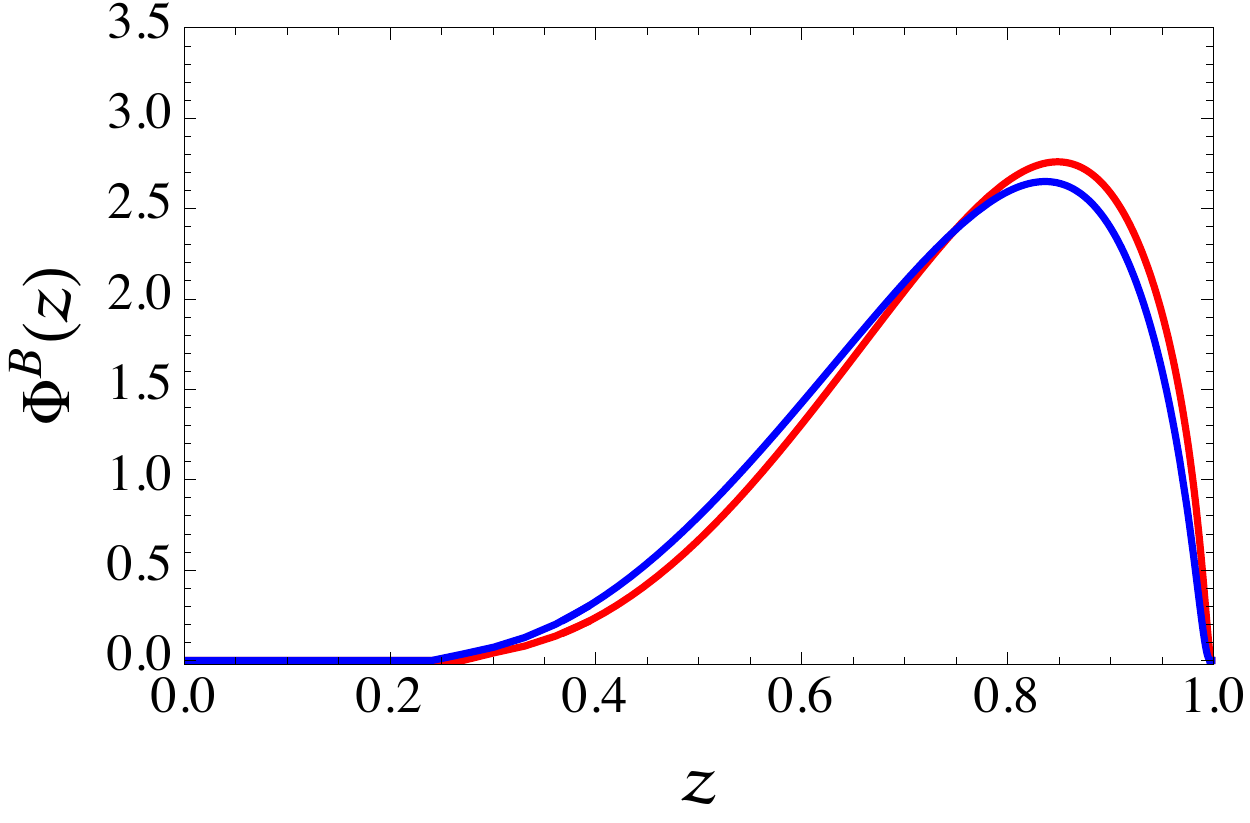}}
\caption{\label{fig:das} \tb{The holographic DAs of $\pi$, $K$, $D_{d,s}$ and $B_{u,s}$ mesons  at $\mu=1\,{\rm GeV}$. The red and blue  curves correspond to the $q-$flavor~($q=u,d$) and $s-$flavor mesons, respectively. } }
\end{center}
\end{figure}

In the following evaluation of \tb{$B_{d,s}\to D_{d,s} P$} decays, the holographic distribution amplitudes (DAs)  are need as hadronic inputs. The DA for pseudoscalar meson parametrizes the operator product expansion of meson-to-vacuum matrix element, which is defined as~\cite{Braun:1989iv}
\begin{eqnarray}
\label{eq:twi2}
\la 0 | \bar q(0)\r_\u\r_5 q(x)|P(p)\ra=
  if_P p_\u \int_0^1 du \, e^{-i up\cdot x} \Phi(u)\,.
\end{eqnarray}
Expanding the hadronic state with the same manner as derivation for the decay constant, we can finally arrive at
\tb{
\begin{eqnarray}
\Phi(z, \u) =  \frac{\sqrt{N_c } }{\pi f_P  }  \int^{|\mathbf{k}_{\bot}| < \mu} \frac{\d^2{\bf k}_{\bot}}{(2\pi)^2}\frac{ \psi(z,{\bf k}_{\bot})}{\sqrt{z\bar{z}}}\frac{\left(\bar{z}m_q+zm_{\bar{q}}\right)}{\sqrt{2}\sqrt{m_0^2-(m_q-m_{\bar{q}})^2}}\,.
\end{eqnarray}
}
in which, the scale $\mu$ with the ultraviolet cut-off on transverse momenta is identified. 

Using the best-fit values of holographic parameters, the DAs of $D_{d,s}$ and $B_{u,s}$ mesons at $\mu=1\,{\rm GeV}$ are plotted in Fig.~\ref{fig:das}, in which the DAs for $\pi^-$ and $K^-$ mesons are also replotted.  With the normalization factor determined by \tb{Eq.~\eqref{eq:NFact}},  all of the DAs can automatically satisfy the normalization condition $\int_{0}^{1}\d z\, \Phi(z)=1$. It can be clearly seen from Fig.~\ref{fig:das} \tb{that the} location where $z$ peaked is close to $1$ and the very small $z\lesssim 0.15$ for $D_{d,s}$ and $z\lesssim 0.5$ for $B_{u,s}$ are strongly suppressed, which imply  the relatively heavier the quark, the larger momentum it can carried, as ones expected.  \tb{The shapes of DAs in Fig.~\ref{fig:das} are generally agree with the ones obtained in, for instance, the LFQM with Gaussian and power-law types WFs~\cite{Hwang:2010hw}~(see Figs.~1 and 2 in Ref.~\cite{Hwang:2010hw} ), the relativistic potential model~\cite{Sun:2016avp}. }

\section{Application in  $B\to D P$ decays}
The decay constants and DAs obtained in the framework of LF holographic QCD given in the last section can be used to evaluate meson decays. In this section,  they will be applied in evaluating  $B_{d,s}\to D_{d,s} P$ ($P=\pi\,,K$) decays.
The effective Hamiltonian responsible for the $B_{d,s}\to D_{d,s} P$ decays, which are induced by the tree-dominated  $b\to c$ transition, can be written as
\begin{equation}\label{Heff}
  {\cal H}_{eff}=\frac{G_{F}}{\sqrt{2}}\sum_{q'=d,s}V_{cb}V^{*}_{uq'}
  \Big\{C_{1}(\mu)Q_{1}(\mu)+C_{2}(\mu)Q_{2}(\mu)\Big\}+h.c.,
\end{equation}
where $G_{F}$ is the Fermi coupling constant, $V_{cb}V^{*}_{uq'}$ is the product of CKM matrix elements~\cite{Cabibbo:1963yz,Kobayashi:1973fv}, $Q_{1,2}$ are local tree four-quark operators, $C_{1,2}(\mu)$ are Wilson coefficients summarize the physical contributions above scale of ${\mu}$ and are calculable with the perturbation theory~\cite{Buchalla:1995vs}. 

\begin{figure}[t]
\begin{center}
\includegraphics[scale=0.12]{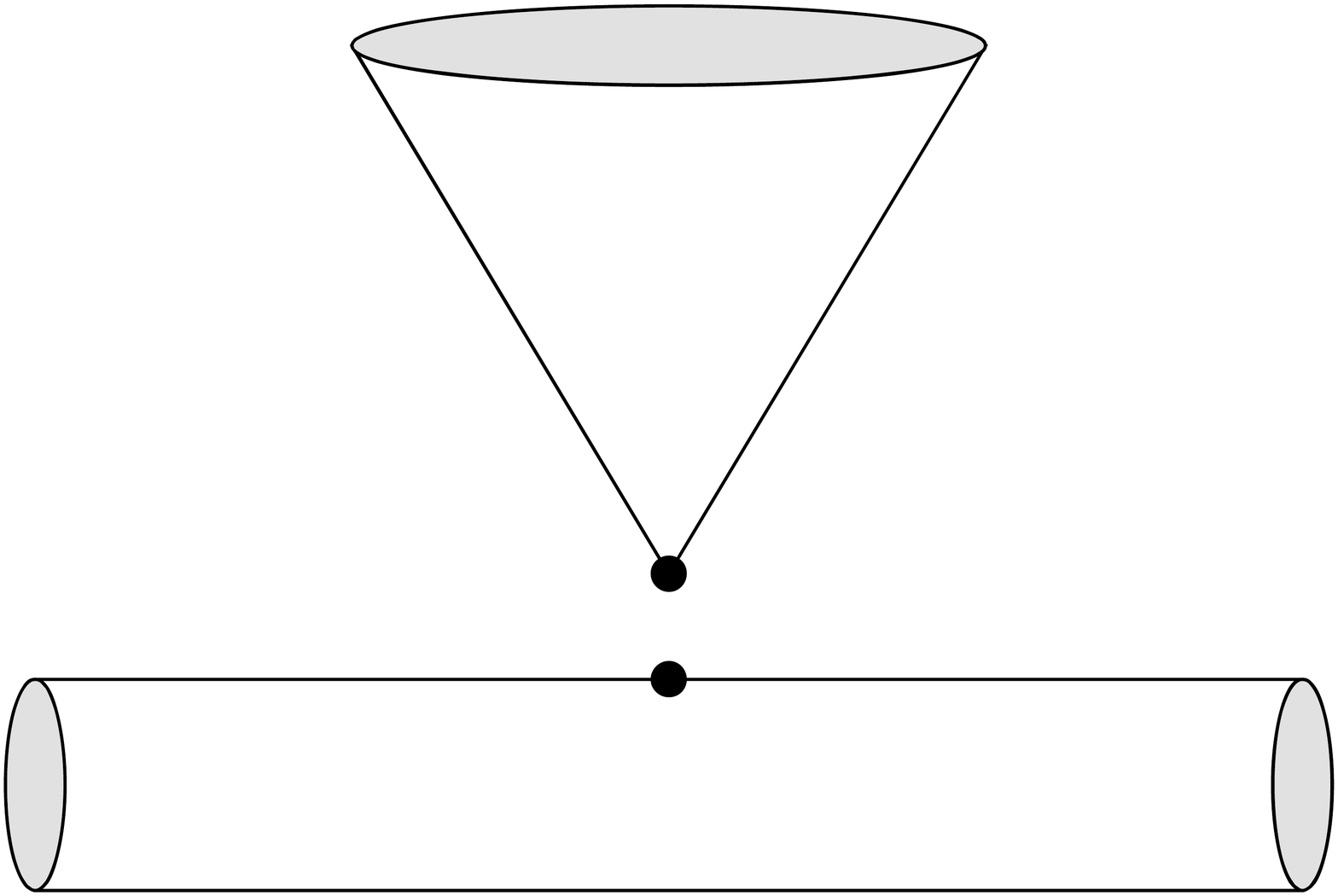}\quad
\caption{\label{fig:t} The leading-order contribution (tree diagram).}
\end{center}
\end{figure}

\begin{figure}[t]
\begin{center}
\subfigure[]{\includegraphics[scale=0.3]{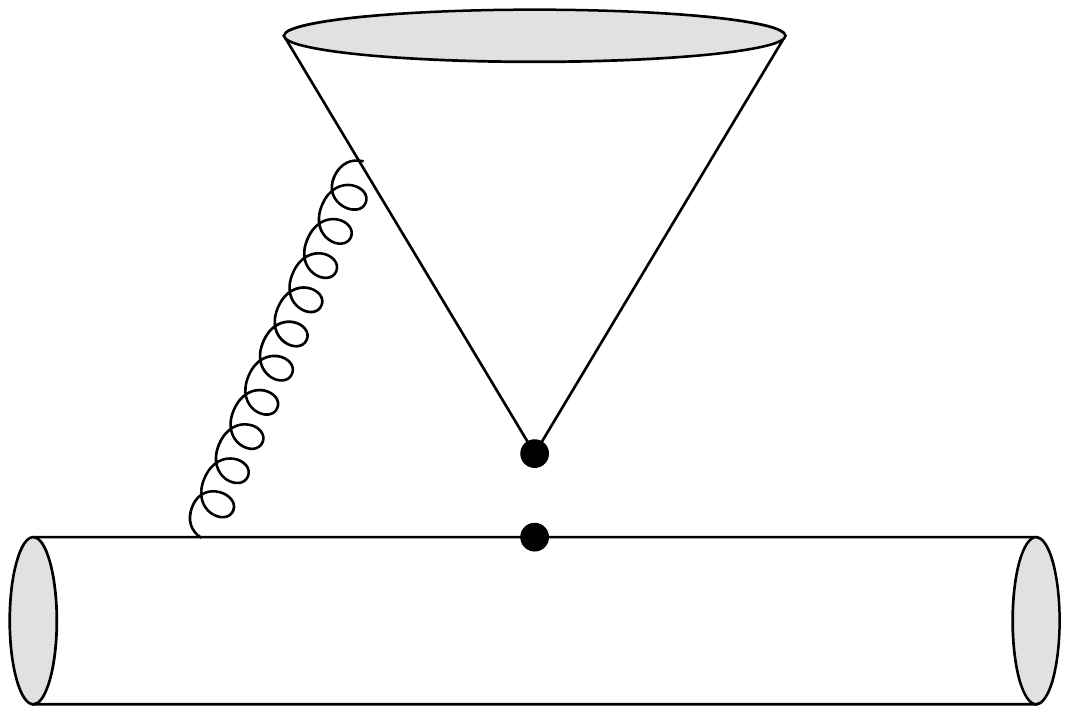}}\quad
\subfigure[]{\includegraphics[scale=0.3]{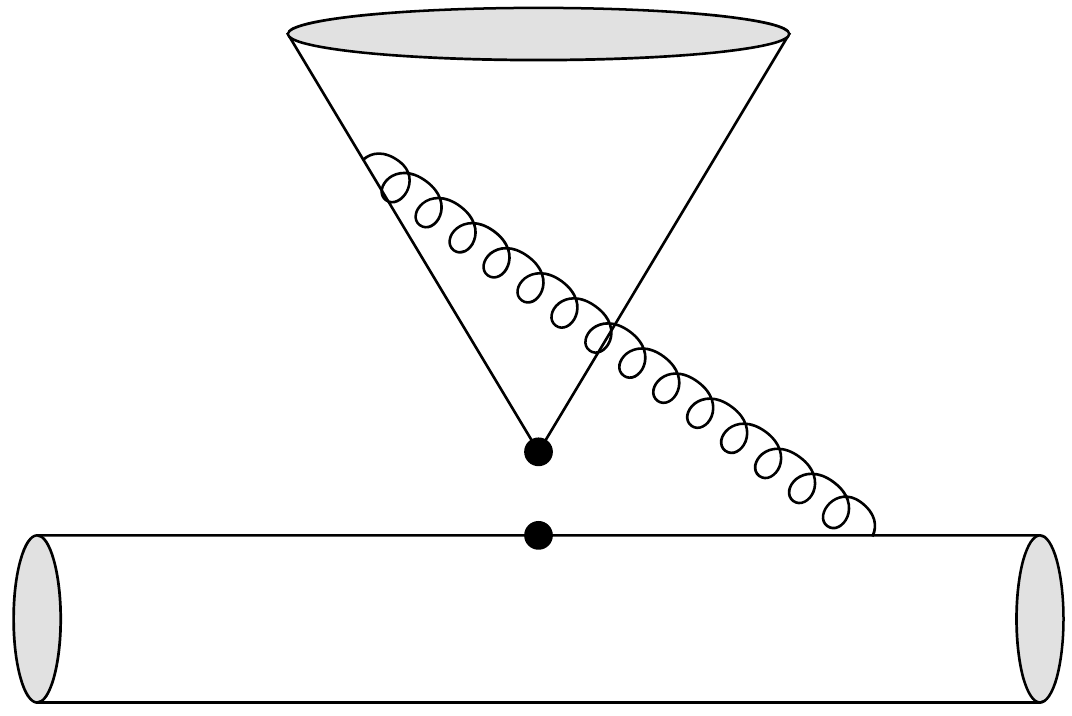}}\quad
\subfigure[]{\includegraphics[scale=0.3]{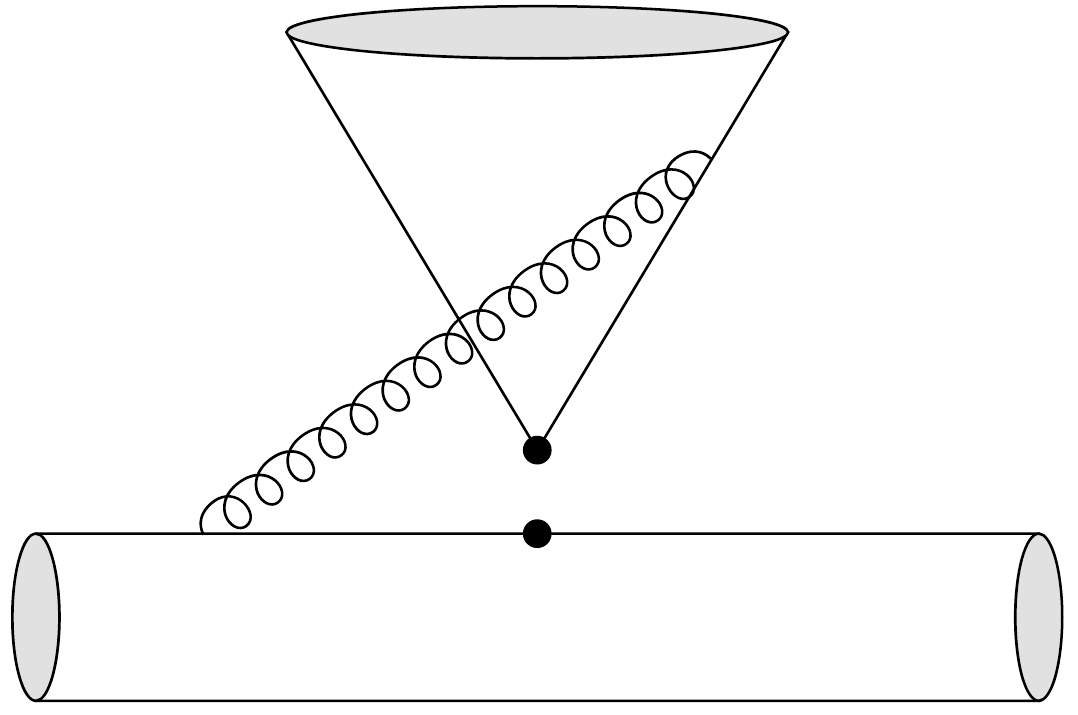}}\quad
\subfigure[]{\includegraphics[scale=0.3]{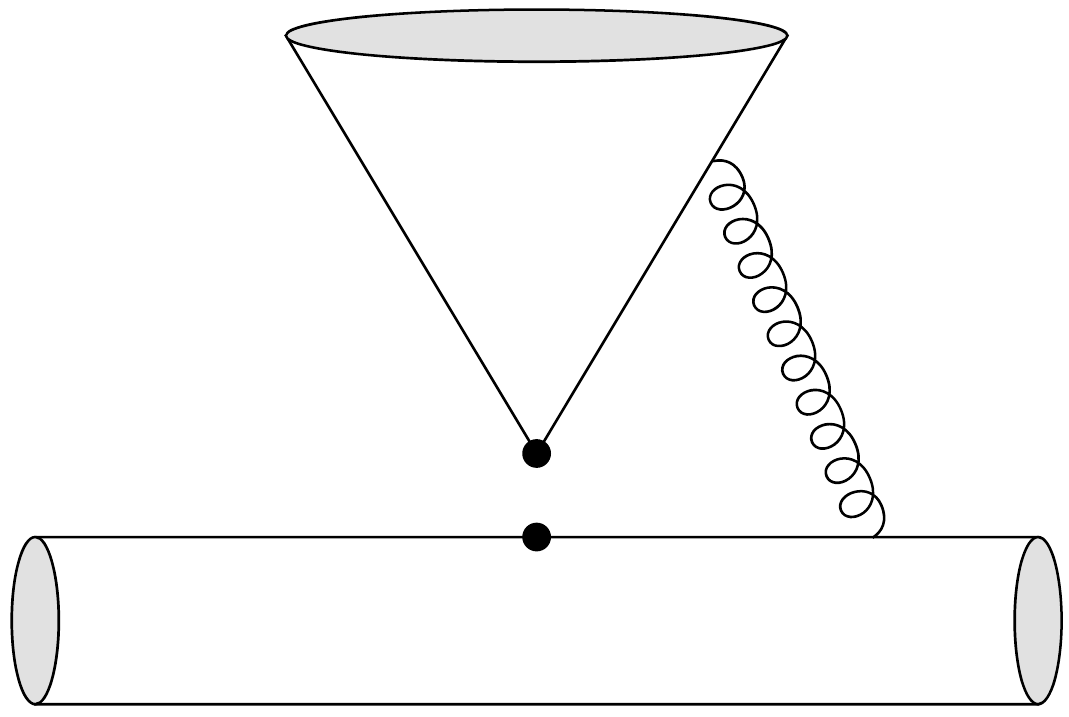}}
\caption{\label{fig:v} The vertex diagrams at the order of $\alpha_s$.}
\end{center}
\end{figure}

\begin{figure}[t]
\begin{center}
\subfigure[]{\includegraphics[scale=0.3]{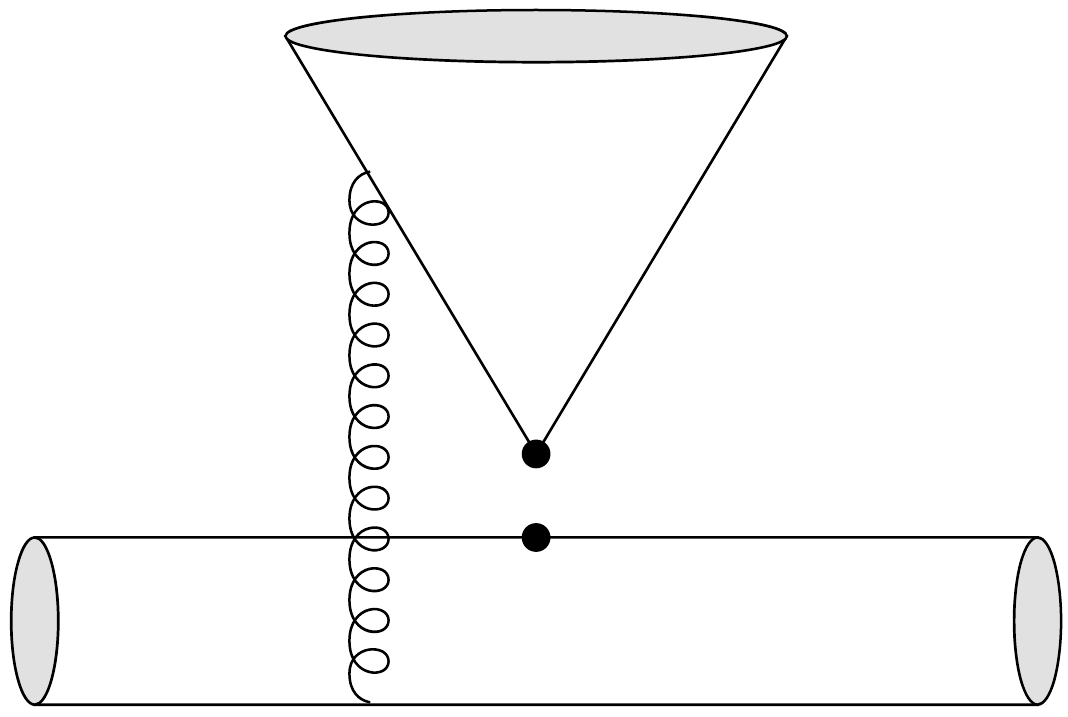}}\quad
\subfigure[]{\includegraphics[scale=0.3]{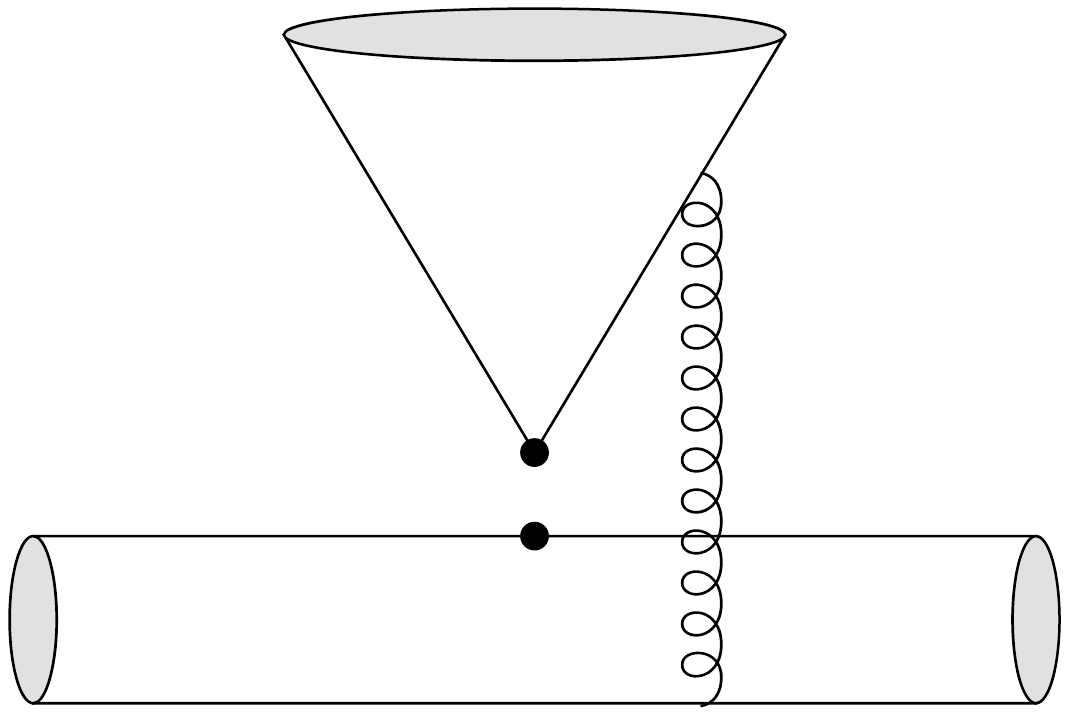}}
\caption{\label{fig:ss} The  spectator scattering  diagrams at the order of $\alpha_s$.}
\end{center}
\end{figure}

\begin{figure}[t]
\begin{center}
\subfigure[]{\includegraphics[scale=0.5]{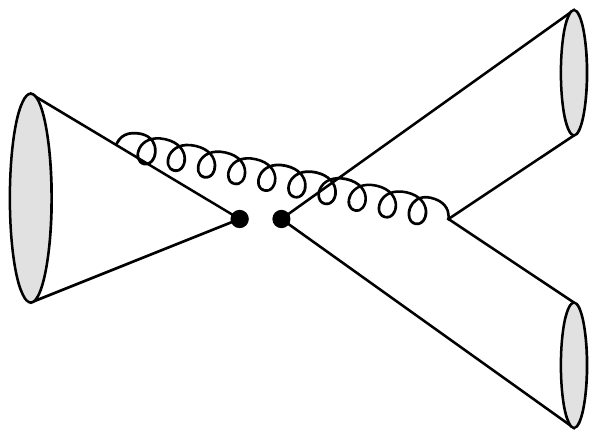}}\quad
\subfigure[]{\includegraphics[scale=0.5]{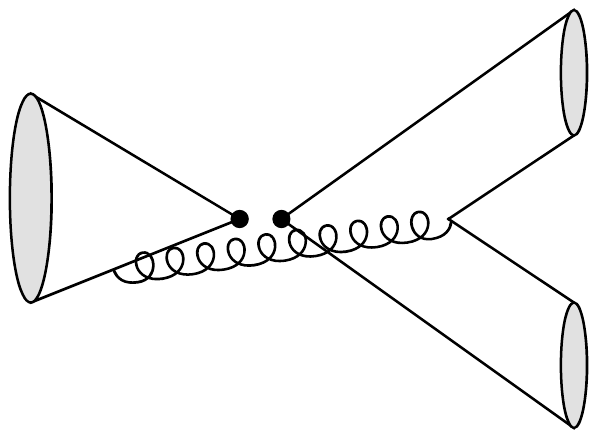}}\quad
\subfigure[]{\includegraphics[scale=0.5]{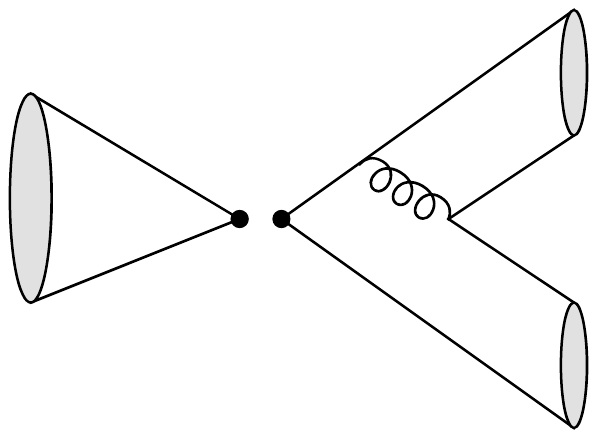}}\quad
\subfigure[]{\includegraphics[scale=0.5]{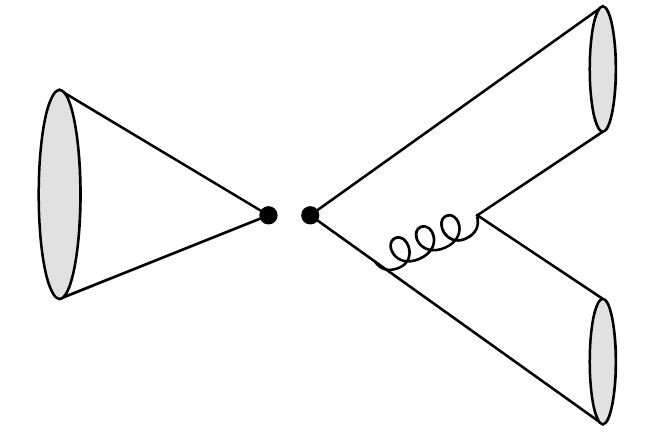}}
\caption{\label{fig:wa} The annihilation  diagrams at the order of $\alpha_s$.}
\end{center}
\end{figure}

In order to obtain the decay amplitudes, the main work is to  accurately calculate the hadronic matrix elements of  local operators in \tb{the} effective Hamiltonian. Following the prescription proposed in Ref.~\cite{Lepage:1980fj}, the hadronic matrix elements for $B\to M_1M_2$ decay can be written as the convolution integrals of the scattering kernel with the DAs of the participating mesons~\cite{Beneke1},
  \begin{eqnarray}\label{eq:fe}
  \langle M_1M_2|Q_i|\bar{B}\rangle &=&\sum_jF_j^{\bar{B}\to M_1}\int dx{\cal T}_{ij}^I(x)\varphi_{M_2}(x)+(M_1\leftrightarrow M_2)\nonumber\\
  &&+\int \d x \d y \d z\,{\cal T}_{i}^{II}(x,\eta,\xi)\,\varphi_{M_1}(x)\,\varphi_{M_2}(\eta)\,\varphi_{B}(\xi)\,,
\label{element}
\end{eqnarray}
where $x\,,\eta\,,\xi$  are the momentum fractions, and the kernels ${\cal T}_{i}^{I,II}(x,\eta,\xi)$ are hard-scattering functions and are perturbatively calculable. 

For the case of $B_{d,s}\to D_{d,s} P$ decays, the first term on the \tb{right-hand-side} of  Eq.~\eqref{eq:fe}  accounts for the tree and vertex amplitudes; the diagrams of tree~(T) and vertex~(V) interaction at NLO are shown in Figs.~\ref{fig:t} and \ref{fig:v}, respectively; the second term vanishes; and the third term accounts for the spectator scattering~(SS) and weak annihilation~(WA) corrections shown by  Figs.~\ref{fig:ss} and \ref{fig:wa}, and is usually dropped in the previous works because it is \tb{power-suppressed~\cite{Beneke2}}.  It should be noted that, some of the power-suppressed corrections, such as the transverse amplitudes for $B\to D^* V$ decay providing $\sim 10\%$ contribution \tb{on the branching fraction}~\cite{Chang:2016eto},  are possibly essential for the relatively accurate \tb{predictions}; moreover, it is also worth to test if the power-suppressed corrections are comparable to the NNLO QCD corrections, which add another positive shift of $2-3\%$ on the amplitude level~\cite{Huber:2016xod}. Therefore, in this work, the SS and WA contributions are reserved, and will be evaluated. In addition, it is arguable if the SS amplitude can be calculated perturbatively~\cite{Beneke2}, nonetheless, the QCD factorization formula, Eq.~\eqref{eq:fe}, is still employed in our estimation.

Applying the QCDF formula, the amplitude of $\bar{B_q}\to D_q P$ decay can be written as
\begin{eqnarray}
  {\cal M}(\bar{B_q}\to D_q P) &=& \frac{G_{F}}{\sqrt{2}}\, V_{cb} V_{uq'}^{\ast}\, A  (\alpha_{1}\,+\beta_1)\,,
   \label{amp}
\end{eqnarray}
with
\begin{eqnarray}
A\equiv if_M F_0^{B\to D}(q^2)(m_B^2-m_D^2)\,,
\end{eqnarray}
where $F_0^{B\to D}$ is the form factor; $q'=d$ and $s$ for $P=\pi$ and $K$, respectively; $\alpha_{1}$ and $\beta_1$ are the effective flavor coefficients. The tree, vertex and  spectator scattering contributions are included in $\alpha_{1}$; the annihilation correction is included in $\beta_1$, which is zero for $\bar{B_d}\to D_d K$ and  $\bar{B_s}\to D_s \pi$ decays at the order of $\alpha_s$. 
These effective coefficients in Eq.~\eqref{amp} are explicitly written as 
 \begin{eqnarray}\label{eq:effc}
\alpha_{1}&=& C_{1}+\frac{1}{N_{c}}\,C_{2}+ C_{2}\,\frac{C_{F}}{N_{c}}\,\frac{{\alpha}_{s}}{4{\pi}}\, \left(\, V_1+\frac{4\pi^2}{N_c}\frac{B}{A}H_1\right)\,,\\
\beta_1&= & C_{1}\,{\alpha}_{s} \pi \frac{C_{F}}{N_{c}^2}\frac{B}{A}A_1\,,
\end{eqnarray}
where $B\equiv if_Bf_Df_P$;
$V_1$, $H_1$, $A_1$ are the vertex, spectator scattering and annihilation functions, respectively, and are obtained by calculating Figs.~\ref{fig:v}, ~\ref{fig:ss} and \ref{fig:wa}. Without the QCD corrections, the naive factorization~(NF) result, {\it i.e.}, the tree contribution, $\alpha_{1}=C_1+C_2/N_c$, can be recovered. 

For the vertex function,  after calculating Fig.~\ref{fig:v}, one can obtain 
\begin{eqnarray}\label{eq:v10}
  V_1 ={\int}_{0}^{1}dx\, {\Phi}_{M}(x)
  \left[ 3\,{\log} \Big( \frac{ m_{b}^{2} }{ {\mu}^{2} } \Big)+ 3\,{\log} \Big( \frac{ m_{c}^{2} }{ {\mu}^{2} } \Big)- 18 +g_0(x)\right]\,,
 \end{eqnarray} 
in which  
\begin{eqnarray}\label{eq:g0}
 g_0(x)&=&
      \frac{c_{a}}{1-c_{a}}\,{\log}(c_{a})-\frac{4\,c_{b}}{1-c_{b}}\,{\log}(c_{b})
      +\frac{c_{d}}{1-c_{d}}\,{\log}(c_{d})-\frac{4\,c_{c}}{1-c_{c}}\,{\log}(c_{c})
      \nonumber \\ &&+
      f(c_{a})-f(c_{b})-f(c_{c})+f(c_{d})
      +2\, {\log}(z_{c}^{2}) \big[ {\log}(c_{a}) -{\log}(c_{b}) \big] \nonumber \\ &&
      -z_{c}\, \Big[ \frac{c_{a}}{(1-c_{a})^{2}}\,{\log}(c_{a})+ \frac{1}{1-c_{a}} \Big]
      -z_{c}^{-1}\,\Big[ \frac{c_{d}}{(1-c_{d})^{2}}\,{\log}(c_{d})+ \frac{1}{1-c_{d}} \Big]\,,
  \end{eqnarray} 
with $z_{c} = m_{c}/m_{b}$, $c_{a} = x\,(1-z_{c}^{2})$, $c_{b} = \bar{x}\,(1-z_{c}^{2})$,  $c_{c} = -c_{a}/z_{c}^{2}$, $c_{d} = -c_{b}/z_{c}^{2}$ and
  \begin{eqnarray}
  f(c)=2{\rm Li}_2(\frac{c-1}{c})-\log^2(c)-\frac{2c}{1-c}\log(c)\,.
  \end{eqnarray}
These results are in agreement with the ones obtained  in previous works, for instance, Ref.~\cite{Beneke2}. In addition, the vertex function for $B\to PP$ decay, which has been given in, for instance, Refs.~\cite{Beneke:2001ev,Beneke:2003zv,du2}, can be recovered from above formulae by taking the limit of  $m_c\to0$.

For the spectator scattering function, after calculating Fig.~\ref{fig:ss}, we  finally obtain
\begin{eqnarray}
H_{1}&=&H_{1}^a+H_{1}^b\,,\\
H_{1}^a&=&\int_0^1 dx\,d \eta\,d\xi\, \Phi_P(x)\,\Phi_D(\eta)\,\Phi_{B}(\xi)\, (1-r_D^2) \,  
\nonumber\\
&&\frac{r_D (\bar{\xi}-\bar{ \eta})+ 2\bar{\xi}-\bar{ \eta}(1+ r_D^2)-\bar{x}(1- r_D^2)}
{[\bar{\xi}\bar{ \eta}(1+ r_D^2)][\bar{\xi}\bar{ \eta}(1+ r_D^2)+\bar{x}(\bar{\xi}-\bar{ \eta})(1- r_D^2)]}\, \\
H_1^b&=&\int_0^1 dx\,d \eta\,d\xi\, \Phi_P(x)\,\Phi_D(\eta)\,\Phi_{B}(\xi)\, (1-r_D^2)\nonumber\\
&& \frac{-r_D (\bar{\xi}-\bar{ \eta})-\bar{\xi}(1+r_D^2)+2\bar{ \eta}r_D^2+x(1-r_D^2) }
{[\bar{\xi}\bar{ \eta}(1+r_D^2)][\bar{\xi}\bar{ \eta}(1+r_D^2)+x(\bar{\xi}-\bar{ \eta})(1-r_D^2)]}\, 
\end{eqnarray}    
where, $r_D=m_D/m_B$; the momentum fractions $x$, $\eta$ and $\xi$ ($\bar{x}\equiv 1-x$, $\bar{\eta}\equiv 1-\eta$ and $\bar{\xi}\equiv 1-\xi$) for the quark (anti-quark) in $P$, $D$ and $B$ mesons are specified, respectively.  The twist-3 DA of $P$ meson doesn't contribute to the spectator scattering amplitude. 

For the annihilation function, after calculating the diagrams in Fig.~\ref{fig:wa}, we finally obtain the full expression for $A_1$, which is given in appendix. From Eqs.~(\ref{eq:a1a}-\ref{eq:a1d}), we find that, the twist-3 DA of $P$  meson in addition to twist-2 also contributes to the annihilation amplitudes due to the un-negligible $m_D$; it is obviously different from the case in $B\to PP$ decays~\cite{Beneke:2001ev,Beneke:2003zv}. However, these contributions are strongly suppressed by both $r_D$ and $r_{\mu_P}$ relative to the contributions of twist-2 part, and therefore, they are neglected in our following numerical evaluation.  With such approximation, the annihilation amplitudes, Eqs.~(\ref{eq:a1a}-\ref{eq:a1d}),  can be finally simplified as 
  \begin{eqnarray}
 A_1&=&A_1^a+A_1^b+A_1^c+A_1^d\,,\\
 A_1^{a}&=&\int_0^1 dx\,d \eta\,d\xi\, \Phi_P (x)\Phi_D(\eta)\,\Phi_{B}(\xi)\,
 \frac{(1+r_D^2)( \xi -\bar{\eta} )-r_b}{ [\bar{\eta}x][ x(\bar{\eta}-\xi)(1-r_D^2)-\xi \bar{\eta} (1+r_D^2)]}\,,\label{eq:a1a}
  \\
 A_1^{b} &=&\int_0^1 dx\,d \eta\,d\xi\, \Phi_P (x)\,\Phi_D(\eta)\,\Phi_{B}(\xi)
 \,\frac{  2\bar{\eta} r_D^2+x( 1-r_D^2)-\bar{\xi} (1+r_D^2)}{[\bar{\eta}x][x(\bar{\eta}-\bar{\xi})(1-r_D^2)-\bar{\xi} \bar{\eta} (1+r_D^2) ]}\,  ,\label{eq:a1b}
  \\
 A_1^{c}&=&-\int_0^1 dx\,d \eta\,d\xi\,  \Phi_P (x)\,\Phi_D(\eta)\,\Phi_{B}(\xi)\,
 \frac{2r_c r_D+r_D^2+x( 1-r_D^2)}{[\bar{\eta}x][x(1-r_D^2)+2\eta\bar{\eta}r_D^2]}\,, \label{eq:a1c}\\
 A_1^{d}&=&\int_0^1 dx\,d \eta\,d\xi\, \Phi_P (x)\, \Phi_D(\eta)\,\Phi_{B}(\xi)\,
 \frac{1}{x\bar{\eta}(1-r_D^2)}\,,\label{eq:a1d}
\end{eqnarray}
where $r_c=m_c/m_B$ and $r_b=m_b/m_B$.

The results of $V_1$ had been presented in previous works, for instance Ref.~\cite{Beneke2}, while the SS and WA amplitudes for $B_q\to D_q P$ decays given above are first presented in this paper. The SS and WA amplitudes for the case of $PP$~($P$ is light meson) final states have been fully evaluated, for instance, in Refs~\cite{Beneke:2001ev,Beneke:2003zv}. Such results of the  twist-2 parts for $B\to P P$ decays, for instance, the first terms in Eq.~(47) and Eq.~(54) in Ref.~\cite{Beneke:2003zv}, can be easily recovered from the formulae given above for  $B_q\to D_q P$ decays by taking the heavy-quark limit and assuming light final states, {\it i.e.}, taking $r_{c,D}=0$, $r_{b}=1$, $\xi=1$~($\bar{\xi}=0$) and $\int\d\xi\Phi_B(\xi)/\bar{\xi}=m_B/\lambda_B$. 

In our numerical evaluation, the decay constants and holographic DAs for the mesons given in the last section are used as hadronic inputs. For  the strong coupling constant, the quark masses and the CKM matrix elements, we take
\begin{eqnarray}
&&\alpha_s(m_W)=0.1181\pm 0.0011\text{~\cite{PDG}}\,;\\
&&m_t^{\rm pole}=174.2\pm1.4\,{\rm GeV}\,,~m_b^{\rm pole}=4.78\pm0.06\,{\rm GeV}\,,~m_c^{\rm pole}=1.67\pm0.07\,{\rm GeV}\text{~\cite{PDG}}\,;\\
&&|V_{cb}|=0.04181^{+0.00028}_{-0.00060}\,,~|V_{ud}|=0.974334^{+0.000064}_{-0.000068}\,,~|V_{us}|=0.22508^{+0.00030}_{-0.00028}~\text{\cite{CKMfitter}}\,. 
\end{eqnarray}
For the $B_d\to D_d$ transition form factors, we adopt the CLN parameterization~\cite{Caprini:1997mu}, with the relevant parameters extracted from the experimental data of exclusive semileptonic $b\to c \ell \bar{\nu}_{\ell}$ decays~\cite{HFAG}, 
\begin{eqnarray}
G(1)|V_{cb}|=(41.57\pm 0.45 \pm0.89)\times 10^{-3}\,,~\rho^2=1.128\pm 0.024\pm0.023\,.
\end{eqnarray}
For the $B_s\to D_s$ transition form factor, on the other hand, we use the results obtained by QCD sum-rule techniques, assuming a polar dependence on $q^2$ that is dominated by the nearest resonance~\cite{Blasi:1993fi,ref:np},
\begin{eqnarray}\label{eq:f0qcds}
F_0^{B_s\to D_s}(0)=0.7\pm0.1\,,~m_{\rm res}=6.8\,{\rm GeV}\,.
\end{eqnarray}
\tb{
In fact, one can also evaluate the form factor by using the improved holographic wavefunction. For instance, adopting the strategy of the covariant LF quark model, one can obtain~\cite{Cheng:1996if,Cheng:2003sm}
\begin{eqnarray}
F_0^{B_s\to D_s}(0)=f^{B_s\to D_s}_+(0)&=&\int_0^1\d z\int\frac{\d^2{\bf k}_{\bot}}{2(2\pi)^3} \frac{ \psi_{B_s}^*(z,{\bf k}_{\bot}) \psi_{D_s}(z,{\bf k}_{\bot})}{z\bar{z}}\nonumber\\
&&\frac{(\bar{z}m_b+zm_{s})(\bar{z}m_c+zm_{s})+{\bf k}_{\bot}^2}{\sqrt{m_{0,B_s}^2-(m_b-m_{s})^2}\sqrt{m_{0,D_s}^2-(m_c-m_{s})^2}}
\,
\end{eqnarray}
at the maximum recoil. Numerically, we obtain $F_0^{B_s\to D_s}(0)=0.81\pm0.01$, which is roughly in agreement with the result of QCD sum-rule, Eq.~\eqref{eq:f0qcds}, within the theoretical uncertainties; but the central value is relatively large. 
In the following evaluations, Eq.~\eqref{eq:f0qcds}  is used as input; the results for the branching ratios with $F_0^{B_s\to D_s}(0)=0.81\pm0.01$ can be naively obtained by multiplying a factor $\sim(0.81/0.7)^2$ if the SS and WA contributions are trivial.
}
Besides, for the other well-determined input parameters, such as the Fermi coupling constant, the lifetimes of $B_{d,s}$ mesons and the masses of mesons, we take the central values given by PDG2016~\cite{PDG}. The theoretical uncertainties in our evaluation are obtained by evaluating separately the uncertainties induced by each input parameter given above and the renormalization scale $\mu={m_b}^{+m_b}_{-m_b/2}$, and then adding them in quadrature.

\begin{table}[t]
\caption{\label{tab:effa} \tb{The  experimental results  and \tb{theoretical predictions} for $|\alpha_1|$ with and without the SS corrections taken into account. }}
\renewcommand*{\arraystretch}{1.1}
\begin{center}\setlength{\tabcolsep}{5pt}
\begin{tabular}{lccccccccc}
\hline\hline
        & T$+$V&T$+$V$+$SS & NLO~\cite{Huber:2016xod}& NNLO~\cite{Huber:2016xod}&Exp.~\cite{Huber:2016xod} \\ \hline
$|\alpha_1(D_d\pi)|$&$1.054^{+0.020}_{-0.018}$ &$1.051^{+0.018}_{-0.016}$
&$1.054^{+0.022}_{-0.020}$&$1.073^{+0.012}_{-0.014}$& $0.89\pm 0.05$ \\
$|\alpha_1(D_d K)|$&$1.053^{+0.019}_{-0.017}$&$1.049^{+0.017}_{-0.016}$ 
&$1.054^{+0.022}_{-0.019}$& $1.070^{+0.010}_{-0.013}$& $0.87 \pm0.06$  \\
$|\alpha_1(D_s \pi)|$&  $1.054^{+0.020}_{-0.018}$ &$1.051^{+0.018}_{-0.017}$
&--- &---& --- \\
$|\alpha_1(D_s K)|$&  $1.053^{+0.019}_{-0.017}$  &$1.050^{+0.017}_{-0.016}$ 
&---&---& --- \\
\hline\hline
\end{tabular}
\end{center}
\end{table}

\begin{table}[t]
\caption{\label{tab:br} \tb{Theoretical predictions for the CP-averaged branching fractions (in units of $10^{-3}$ for $\bar{B}_{d,s}\to D_{d,s}^+\pi^-$ and $10^{-4}$ for $\bar{B}_{d,s}\to D_{d,s}^+K^-$  decays) .}}
\renewcommand*{\arraystretch}{1.1}
\begin{center}\setlength{\tabcolsep}{5pt}
\begin{tabular}{lccccccccc}
\hline\hline
            & $T+V$ &$T+V+SS+WA$ & Ref.~\cite{Huber:2016xod}&Exp.~\cite{PDG} \\ \hline
$\bar{B}_d\to D^+\pi^-$&$3.606^{+0.422}_{-0.408}$ &$3.582^{+0.415}_{-0.404}$ 
&$3.93^{+0.43}_{-0.42}$ &$2.52\pm0.13$\\
$\bar{B}_d\to D^+K^-$&$2.731^{+0.319}_{-0.308}$ &$2.712^{+0.313}_{-0.305}$ 
&$3.01^{+0.32}_{-0.31}$   &$1.86\pm0.20$  \\
$\bar{B}_s\to D_s^+\pi^-$&  $4.741^{+1.463}_{-1.274}$&  $4.717^{+1.457}_{-1.270}$ 
&$4.39^{+1.36}_{-1.19}$&$3.00\pm0.23$\\
$\bar{B}_s\to D_s^+K^-$&  $3.591^{+1.108}_{-0.965}$ &  $3.575^{+1.104}_{-0.962}$  
&$3.34^{+1.04}_{-0.90}$ &$2.27\pm0.19$\\
\hline\hline
\end{tabular}
\end{center}
\end{table}

Using the theoretical formulae and inputs given above, we then present our numerical results. Our theoretical predictions for $|\alpha_1|$ with and without the SS corrections 
taken into account are summarized in Table~\ref{tab:effa}, in which, the theoretical predictions~\cite{Huber:2016xod} with vertex corrections at NLO and NNLO level and the results extracted from experimental data~\cite{Huber:2016xod} are also listed for comparison. It can be found that: (i) our results  for $|\alpha_1|$ obtained by using holographic DAs are in consistence with the ones in Ref~\cite{Huber:2016xod} obtained by using lattice inputs~\cite{Arthur:2010xf}; while, both of them are a little larger than the data. (ii) Such possible tension is hardly to be  moderated by the SS correction, which is suppressed by not only $B/A$ but also small $C_2$~(see Eq.\eqref{eq:effc}), and therefore, is too small compared with  the tree amplitude. Numerically,   the SS corrections  present negative shifts of \tb{about $\sim 0.3\%$ on the amplitude level.} 

The WA amplitude is proportional to the large Wilson coefficient $C_1$, and thus, is expected to present a large correction relative to the SS amplitude. However, numerically,  we find
\tb{
 \begin{eqnarray}\label{eq:beta1}
\beta_1(D_d\pi)\times 10^{3}=-0.43^{+0.09}_{-0.14}\,,\qquad
\beta_1(D_s K)\times 10^{3}=0.47^{+0.17}_{-0.11}\,,
  \end{eqnarray}
}
which are very small  and at the same level as  SS contribution. It is mainly caused by that: (i) the WA amplitude is power-suppressed as the SS amplitude. (ii) For the case of light $PP$ final states, the negative $A_1^c$ and the positive $A_1^d$~\tb{(factorizable WA amplitudes)}  cancel entirely with each other, {\it i.e.}, $A_1^c(PP)+A_1^d(PP)=0$, which can be clearly seen from Eqs.~\eqref{eq:a1c} and \eqref{eq:a1d} \tb{or the relevant formulae given in Ref.~\cite{Beneke:2003zv}; therefore,} the WA contribution comes only from positive $A_1^a(PP)+A_1^b(PP)$~\tb{(nonfactorizable WA amplitudes)}. However, for the case of $DP$ final states, because of the corrections induced by the masses of $c$ quark and $D$ meson, $A_1^c(DP)+A_1^d(DP)$ is nonzero and present a relatively  large negative contribution, which further cancels with $A_1^a(DP)+A_1^b(DP)$; as a result, the ``residual"~(total) WA contribution, $A_1$, is very small.  \tb{Taking $B_d\to D_d\pi$ as an example,  we find  $A_1^a+A_1^b\sim 7.6$ and $A_1^c+A_1^d\sim -9.0$ ( the later becomes  zero for the case of two light final states as just mentioned), which results in a very small $A_1\sim -1.4$. Therefore, we would like to emphasize that the small WA contributions to the $B_{d}\to D_{d}\pi$ and $B_{s}\to D_{s}K$ decays are caused by not only the power-suppression but also the significant cancelation between the factorizable and nonfactorizable WA amplitudes. } 

Finally, we present our theoretical \tb{results} for  the CP-averaged branching fractions of $B_{d,s}\to D_{d,s}P$ decays in Table~\ref{tab:br}, in which the experimental data~\cite{PDG} and the theoretical results based on NNLO evaluation~\tb{for the vertex amplitudes}~\cite{Huber:2016xod} are also listed for comparison.  It can be found that the branching fractions of $B_{d,s}\to D_{d,s}P$ decays can only be reduced about \tb{$ 0.7\%\,$ by the}  power-suppressed SS and WA corrections, which is  trivial numerically, and therefore, confirms the analysis based on the power counting rules~\cite{Beneke2}. The central values of the theoretical predictions are still much larger than the data. Such deviation will be further enlarged when the NNLO contribution is taken into account~\cite{Huber:2016xod}. 


\section{Summary}
In this article we have evaluated the decay constant and distribution amplitude for the heavy-light pseudoscalar mesons by using the helicity-improved light-front holographic wavefunction. Then, 
we have further applied these results to evaluate the $B_{d,s}\to D_{d,s}P$~($P$ is light pseudoscalar meson) decays by using QCD factorization approach, in which, besides of the tree and vertex amplitudes, the 
spectator scattering and weak annihilation amplitudes are also taken into account. Our main findings are summarized as follows:
\begin{itemize}
 \item Under the constraints from the decay constants \tb{and masses of} the heavy-light pseudoscalar mesons, we have performed the $\chi^2$-analyses for the holographic parameters $\k$ and the effective quark masses by \tb{using an improved holographic WF}.
 The $\chi^2$-fitting results  are \tb{roughly} in a consistence with the ones obtained by fitting the Regge trajectory \tb{except for some tensions in $D_{(s)}$ system}.
 With the best-fitting values of holographic parameters, we have also predicted the holographic DAs of heavy-light mesons.
  \item With the obtained decay constants and DAs,  we further show that they can be used in the evaluations of $B$ meson decays. The effective coefficients and the branching fractions of $B_{d,s}\to D_{d,s}P$ decays are evaluated in this paper. It is found that the power-suppressed spectator scattering contribution presents negative shifts of about \tb{$0.3\%$} on the amplitude level. Moreover, \tb{the annihilation contribution is also very small because of the significant cancellation between the factorizable  and nonfactorizable WA amplitudes.  }
Such \tb{findings confirm} the previous analysis for the SS and WA corrections based on the power counting rules.
Numerically, the QCDF predictions for the branching fractions of $B_{d,s}\to D_{d,s}P$ decays can be reduced by about \tb{$0.7\%$ by} the power-suppressed corrections; while, \tb{they} are still larger than the data.  
 \end{itemize}

\begin{appendix}
\section*{Appendix: the annihilation amplitude $A_1$}
The full expression for $A_1$:
 \begin{eqnarray}
 A_1&=&A_1^a+A_1^b+A_1^c+A_1^d\,,\\
 \label{eq:a1a}
 A_1^{a}&=&\int_0^1 dx\,d \eta\,d\xi\, \Phi_D(\eta)\,\Phi_{B}(\xi)\,
 \frac{1}{ [\bar{\eta}x(1-r_D^2)][ x(\bar{\eta}-\xi)(1-r_D^2)-\xi \bar{\eta} (1+r_D^2)]}\,\nonumber\\
&&\Big\{ 
(1-r_D^2) \Big[(1+r_D^2)( \xi -\bar{\eta} )-r_b\Big] \Phi_P(x) \nonumber\\
&&+r_D r_{\mu_P} \Big[ \bar{\eta} (1+r_D^2)+4 r_b -2\xi +x (1-r_D^2)\Big] \phi_P(x) 
\Big\}\,,
  \\
 \label{eq:a1b}
 A_1^{b} &=&\int_0^1 dx\,d \eta\,d\xi\, \Phi_D(\eta)\,\Phi_{B}(\xi)
 \,\frac{1}{[\bar{\eta}x(1-r_D^2)][x(\bar{\eta}-\bar{\xi})(1-r_D^2)-\bar{\xi} \bar{\eta} (1+r_D^2) ]}\,  \nonumber\\
&&\Big\{
(1-r_D^2) \Big[ 2\bar{\eta} r_D^2+x( 1-r_D^2)-\bar{\xi} (1+r_D^2)\Big]\Phi_P(x)   \nonumber\\
&&- r_D r_{\mu_P} \Big[ \bar{\eta} (1+r_D^2)+x( 1-r_D^2)-2\bar{\xi}\Big]  \phi_P(x)
\Big\}\,,
  \\
 \label{eq:a1c}
 A_1^{c}&=&\int_0^1 dx\,d \eta\,d\xi\, \Phi_D(\eta)\,\Phi_{B}(\xi)\,
 \frac{1}{[\bar{\eta}x(1-r_D^2)][x(1-r_D^2)+2\eta\bar{\eta}r_D^2]}\, \nonumber\\
&&\Big\{-(1-r_D^2) \Big[2r_c r_D+r_D^2+x( 1-r_D^2)\Big]\Phi_P(x)\nonumber\\
&&+r_{\mu_P} \Big[ r_c (1+r_D^2)+2 r_D  \big((1+r_D^2)+x(1-r_D^2)\big) \Big] \phi_P(x)
\Big\}\,,  \\
 \label{eq:a1d}
 A_1^{d}&=&\int_0^1 dx\,d \eta\,d\xi\, \Phi_D(\eta)\,\Phi_{B}(\xi)\,
 \frac{1}{[\bar{\eta}x(1-r_D^2)]\,[\bar{\eta}(1-r_D^2)]}\, \nonumber\\
&& \Big\{ \bar{\eta} (1-r_D^2) \Phi_P (x)- 2r_D r_{\mu_P} [ (1-r_D^2)+\bar{\eta}(1+r_D^2)]\phi_P(x)  
 \Big\}\,,
\end{eqnarray}
where $r_c=m_c/m_B$, $r_b=m_b/m_B$, $r_D=m_D/m_B$ and $r_{\mu_P}=\frac{m_P^2}{m_B(m_q+m_{\bar{q}})}$~($m_q$ and $m_{\bar{q}}$ are the quark masses in $P$ meson);  $\Phi_P (x)$ and $\phi_P(x)$ are the twist-2 and -3 DAs of $P$ meson, respectively.

\end{appendix}
\section*{Acknowledgements}
We thank Prof. Stanley J. Brodsky at SLAC for helpful discussions. 
This work is supported by the National Natural Science Foundation of China (Grant No. 11475055
), the Foundation for the Author of National Excellent Doctoral Dissertation of China (Grant No. 201317), the Program for Science and Technology Innovation Talents in Universities of Henan Province (Grant No. 14HASTIT036), the Excellent Youth Foundation of HNNU and the CSC (Grant No. 201508410213).



\begin{thebibliography}{99}

\bibitem{deTeramond:2005su}
  G.~F.~de Teramond and S.~J.~Brodsky,
  Phys.\ Rev.\ Lett.\  {\bf 94} (2005) 201601.

\bibitem{Brodsky:2006uqa}
  S.~J.~Brodsky and G.~F.~de Teramond,
  Phys.\ Rev.\ Lett.\  {\bf 96} (2006) 201601.

\bibitem{Brodsky:2007hb}
  S.~J.~Brodsky and G.~F.~de Teramond,
  Phys.\ Rev.\ D {\bf 77} (2008) 056007.

\bibitem{Brodsky:2008pg}
  S.~J.~Brodsky and G.~F.~de Teramond,
  Subnucl.\ Ser.\  {\bf 45} (2009) 139.
  
\bibitem{Brodsky:2008pf}
  S.~J.~Brodsky and G.~F.~de Teramond,
  Phys.\ Rev.\ D {\bf 78} (2008) 025032.

\bibitem{deTeramond:2008ht}
  G.~F.~de Teramond and S.~J.~Brodsky,
  Phys.\ Rev.\ Lett.\  {\bf 102} (2009) 081601.

\bibitem{deTeramond:2010ge}
  G.~F.~de Teramond and S.~J.~Brodsky,
  AIP Conf.\ Proc.\  {\bf 1296} (2010) 128. 
    
\bibitem{Brodsky:2014yha}
  S.~J.~Brodsky, G.~F.~de Teramond, H.~G.~Dosch and J.~Erlich,
  Phys.\ Rept.\  {\bf 584} (2015) 1.  


\bibitem{deTeramond:2014asa}
  G.~F.~de Teramond, H.~G.~Dosch and S.~J.~Brodsky,
  Phys.\ Rev.\ D {\bf 91} (2015) no.4,  045040.
  
  \bibitem{Dosch:2015nwa}
  H.~G.~Dosch, G.~F.~de Teramond and S.~J.~Brodsky,
  Phys.\ Rev.\ D {\bf 91} (2015) no.8,  085016

\bibitem{Brodsky:2016yod}
  S.~J.~Brodsky, G.~F.~de TšŠramond, H.~G.~Dosch and C.~LorcšŠ,
  Phys.\ Lett.\ B {\bf 759} (2016) 171.

\bibitem{Brodsky:2016rvj}
  S.~J.~Brodsky, G.~F.~de TšŠramond, H.~G.~Dosch and C.~LorcšŠ,
  Int.\ J.\ Mod.\ Phys.\ A {\bf 31} (2016) no.19,  1630029.
  
\bibitem{Dosch:2016zdv}
  H.~G.~Dosch, G.~F.~de Teramond and S.~J.~Brodsky,
  arXiv:1612.02370 [hep-ph].
  
 \bibitem{Branz:2010ub}
  T.~Branz, T.~Gutsche, V.~E.~Lyubovitskij, I.~Schmidt and A.~Vega,
  Phys.\ Rev.\ D {\bf 82} (2010) 074022.

\bibitem{Brodsky:2011yv}
  S.~J.~Brodsky, F.~G.~Cao and G.~F.~de Teramond,
  Phys.\ Rev.\ D {\bf 84} (2011) 033001.
  
\bibitem{Hwang:2012xf}
  C.~W.~Hwang,
  Phys.\ Rev.\ D {\bf 86} (2012) 014005.
  
\bibitem{Braga:2015jca}
  N.~R.~F.~Braga, M.~A.~Martin Contreras and S.~Diles,
  Phys.\ Lett.\ B {\bf 763} (2016) 203.

\bibitem{Vega:2009zb}
  A.~Vega, I.~Schmidt, T.~Branz, T.~Gutsche and V.~E.~Lyubovitskij,
  Phys.\ Rev.\ D {\bf 80} (2009) 055014.
  
 \bibitem{Lyubovitskij:2010zz}
\tb{  V.~Lyubovitskij, T.~Branz, T.~Gutsche, I.~Schmidt and A.~Vega,
  PoS LC {\bf 2010} (2010) 030.  
}
   
\bibitem{Vega:2011ck}
\tb{  A.~Vega, I.~Schmidt, T.~Gutsche and V.~E.~Lyubovitskij,
  AIP Conf.\ Proc.\  {\bf 1432} (2012) 253.
}


\bibitem{Teramond:GHP2009}
G. F. de Teramond, 
Third Workshop of the APS Topical Group in Hadron Physics ‘‘GHP2009’’.


 \bibitem{Gutsche:2014oua}
\tb{  T.~Gutsche, V.~E.~Lyubovitskij, I.~Schmidt and A.~Vega,
  Phys.\ Rev.\ D {\bf 90} (2014) no.9,  096007.  
}
  
\bibitem{Swarnkar:2015osa}
  R.~Swarnkar and D.~Chakrabarti,
  Phys.\ Rev.\ D {\bf 92} (2015) no.7,  074023.

\bibitem{Dosch:1996ss}
  H.~G.~Dosch, T.~Gousset, G.~Kulzinger and H.~J.~Pirner,
  Phys.\ Rev.\ D {\bf 55} (1997) 2602.

\bibitem{Forshaw:2012im}
  J.~R.~Forshaw and R.~Sandapen,
  Phys.\ Rev.\ Lett.\  {\bf 109} (2012) 081601.

\bibitem{Ahmady:2012dy}
  M.~Ahmady and R.~Sandapen,
  Phys.\ Rev.\ D {\bf 87} (2013) no.5,  054013.

\bibitem{Ahmady:2013cva}
  M.~Ahmady and R.~Sandapen,
  Phys.\ Rev.\ D {\bf 88} (2013) 014042.

\bibitem{Ahmady:2013cga}
  M.~Ahmady, R.~Campbell, S.~Lord and R.~Sandapen,
  Phys.\ Rev.\ D {\bf 88} (2013) no.7,  074031.

\bibitem{Ahmady:2014sva}
  M.~Ahmady, R.~Campbell, S.~Lord and R.~Sandapen,
  Phys.\ Rev.\ D {\bf 89} (2014) no.7,  074021.

\bibitem{Ahmady:2014cpa}
  M.~R.~Ahmady, S.~Lord and R.~Sandapen,
  Phys.\ Rev.\ D {\bf 90} (2014) no.7,  074010.

\bibitem{Ahmady:2015gva}
  M.~Ahmady, S.~Lord and R.~Sandapen,
  PoS DIS {\bf 2015} (2015) 160.

\bibitem{Ahmady:2015yea}
  M.~Ahmady, S.~Lord and R.~Sandapen,
  Nucl.\ Part.\ Phys.\ Proc.\  {\bf 270-272} (2016) 160.


\bibitem{Ahmady:2016ufq}
  M.~Ahmady, F.~Chishtie and R.~Sandapen,
  arXiv:1609.07024 [hep-ph].
  
  \bibitem{Chang:2017}
  Q.~Chang, S.~J.~Brodsky and X.~Q.~Li,
  arXiv:1612.05298 [hep-ph].
  
\bibitem{Zhou:2015jba}
  S.~H.~Zhou, Y.~B.~Wei, Q.~Qin, Y.~Li, F.~S.~Yu and C.~D.~Lu,
  Phys.\ Rev.\ D {\bf 92} (2015) no.9,  094016.
  
  \bibitem{Lu:2016uwv}
  C.~D.~Lu and S.~H.~Zhou,
  arXiv:1605.08546 [hep-ph].

 \bibitem{Chua:2007qw}
  C.~K.~Chua and W.~S.~Hou,
  Phys.\ Rev.\ D {\bf 77} (2008) 116001.

  
  \bibitem{Beneke2}
  M.~Beneke, G.~Buchalla, M.~Neubert and C.~T.~Sachrajda,
  Nucl.\ Phys.\ B {\bf 591} (2000) 313.
  
\bibitem{Huber:2016xod}
  T.~Huber, S.~Kränkl and X.~Q.~Li,
  JHEP {\bf 1609} (2016) 112. 
      
\bibitem{Zou:2009zza}
  H.~Zou, R.~H.~Li, X.~X.~Wang and C.~D.~Lu,
  J.\ Phys.\ G {\bf 37} (2010) 015002.
  
\bibitem{Rui:2011qc}
  Z.~Rui, Z.~T.~Zou and C.~D.~Lu,
  Phys.\ Rev.\ D {\bf 86} (2012) 074008.

 \bibitem{Melosh:1974cu}
  H.~J.~Melosh,
  Phys.\ Rev.\ D {\bf 9} (1974) 1095.

 \bibitem{Jaus:1989au}
  W.~Jaus,
  Phys.\ Rev.\ D {\bf 41} (1990) 3394.
  
\bibitem{Hwang:2010hw}
  C.~W.~Hwang,
  Phys.\ Rev.\ D {\bf 81} (2010) 114024. 
    
  \bibitem{Brodsky:1997de}
  S.~J.~Brodsky, H.~C.~Pauli and S.~S.~Pinsky,
  Phys.\ Rept.\  {\bf 301} (1998) 299.
  
\bibitem{Lepage:1980fj}
  G.~P.~Lepage and S.~J.~Brodsky,
  Phys.\ Rev.\ D {\bf 22} (1980) 2157.
  
\bibitem{Deur:2016opc}
  A.~Deur, S.~J.~Brodsky and G.~F.~de Teramond,
  arXiv:1608.04933 [hep-ph].

\bibitem{Prosperi:2006hx}
  G.~M.~Prosperi, M.~Raciti and C.~Simolo,
  Prog.\ Part.\ Nucl.\ Phys.\  {\bf 58} (2007) 387.
  


\bibitem{Ahmady:2016ujw}
  M.~Ahmady, R.~Sandapen and N.~Sharma,
  Phys.\ Rev.\ D {\bf 94} (2016) no.7,  074018.  
  
 \bibitem{PDG}
  C.~Patrignani {\it et al.} [Particle Data Group],
  Chin.\ Phys.\ C {\bf 40} (2016) no.10,  100001.
  
\bibitem{Sergeenko:1994ck}
\tb{  M.~N.~Sergeenko,
  Z.\ Phys.\ C {\bf 64} (1994) 315.
 }
  
 \bibitem{Gershtein:2006ng}
\tb{  S.~S.~Gershtein, A.~K.~Likhoded and A.~V.~Luchinsky,
  Phys.\ Rev.\ D {\bf 74} (2006) 016002.
}
 
 \bibitem{Gutsche:2012ez}
\tb{  T.~Gutsche, V.~E.~Lyubovitskij, I.~Schmidt and A.~Vega,
  Phys.\ Rev.\ D {\bf 87} (2013) no.5,  056001.
  }
 
\bibitem{Badalian:2004xv}
\tb{  A.~M.~Badalian, A.~I.~Veselov and B.~L.~G.~Bakker,
  Phys.\ Rev.\ D {\bf 70} (2004) 016007.
}
  
\bibitem{Ebert:2009ub}
\tb{  D.~Ebert, R.~N.~Faustov and V.~O.~Galkin,
  Phys.\ Rev.\ D {\bf 79} (2009) 114029.
}
  
  \bibitem{Aoki:2016frl}
  S.~Aoki {\it et al.},
  arXiv:1607.00299 [hep-lat].
  
\bibitem{Narison:2015nxh}
  S.~Narison,
  Nucl.\ Part.\ Phys.\ Proc.\  {\bf 270-272} (2016) 143.
 
  \bibitem{Hwang:2009qz}
  C.~W.~Hwang,
  Phys.\ Rev.\ D {\bf 81} (2010) 054022.
   
  \bibitem{Braun:1989iv}
  V.~M.~Braun and I.~E.~Filyanov,
  ``Conformal Invariance and Pion Wave Functions of Nonleading Twist,''
  Z.\ Phys.\ C {\bf 48} (1990) 239.
  
\bibitem{Sun:2016avp}
  H.~K.~Sun and M.~Z.~Yang,
  arXiv:1609.08958 [hep-ph].
     
\bibitem{Cabibbo:1963yz}
  N.~Cabibbo,
  Phys.\ Rev.\ Lett.\  {\bf 10} (1963) 531.
    
\bibitem{Kobayashi:1973fv}
  M.~Kobayashi and T.~Maskawa,
  Prog.\ Theor.\ Phys.\  {\bf 49} (1973) 652.
  
    \bibitem{Buchalla:1995vs}
  G.~Buchalla, A.~J.~Buras and M.~E.~Lautenbacher,
  Rev.\ Mod.\ Phys.\  {\bf 68} (1996) 1125.
  
  

\bibitem{Beneke1}
  M.~Beneke, G.~Buchalla, M.~Neubert and C.~T.~Sachrajda,
  Phys.\ Rev.\ Lett.\  {\bf 83} (1999) 1914.



\bibitem{Chang:2016eto}
  Q.~Chang, L.~X.~Chen, Y.~Y.~Zhang, J.~F.~Sun and Y.~L.~Yang,
  Eur.\ Phys.\ J.\ C {\bf 76} (2016) no.10,  523.



\bibitem{Beneke:2001ev}
  M.~Beneke, G.~Buchalla, M.~Neubert and C.~T.~Sachrajda,
  Nucl.\ Phys.\ B {\bf 606} (2001) 245.
  
\bibitem{Beneke:2003zv}
  M.~Beneke and M.~Neubert,
  Nucl.\ Phys.\ B {\bf 675} (2003) 333
  
\bibitem{du2}
  J.~f.~Sun, G.~h.~Zhu and D.~s.~Du,
  Phys.\ Rev.\ D {\bf 68} (2003) 054003.



\bibitem{CKMfitter}
  J.~Charles {\it et al.} [CKMfitter Group Collaboration],
  Eur.\ Phys.\ J.\ C {\bf 41} (2005) 1
  [hep-ph/0406184], updated results and plots available at: http://ckmfitter.in2p3.fr.

\bibitem{Caprini:1997mu}
  I.~Caprini, L.~Lellouch and M.~Neubert,
  Nucl.\ Phys.\ B {\bf 530} (1998) 153.

 \bibitem{HFAG}
  Y.~Amhis {\it et al.} [Heavy Flavor Averaging Group (HFAG) Collaboration],
  arXiv:1412.7515 [hep-ex], and online updates at http://www.slac.stanford.edu/xorg/hfag/.
  
  \bibitem{Blasi:1993fi}
  P.~Blasi, P.~Colangelo, G.~Nardulli and N.~Paver,
  Phys.\ Rev.\ D {\bf 49} (1994) 238.

  \bibitem{ref:np}  
 P. Colangelo, G. Nardulli and N. Paver, 
 Z. Phys. C {\bf 57} (1993) 43. 
 
 \bibitem{Cheng:1996if}
\tb{  H.~Y.~Cheng, C.~Y.~Cheung and C.~W.~Hwang,
  Phys.\ Rev.\ D {\bf 55} (1997) 1559.}

\bibitem{Cheng:2003sm}
 \tb{ H.~Y.~Cheng, C.~K.~Chua and C.~W.~Hwang,
  Phys.\ Rev.\ D {\bf 69} (2004) 074025.}

 \bibitem{Arthur:2010xf}
  R.~Arthur, P.~A.~Boyle, D.~Brommel, M.~A.~Donnellan, J.~M.~Flynn, A.~Juttner, T.~D.~Rae and C.~T.~C.~Sachrajda,
    Phys.\ Rev.\ D {\bf 83} (2011) 074505.
   
\end{thebibliography}
\end{document}